\documentclass[a4paper,showpacs,preprintnumbers,superscriptaddress,aps,prd,nofootinbib,floatfix,11pt]{revtex4-1}
\usepackage{graphicx}  
\usepackage{dcolumn}
\usepackage{bm}
\usepackage{amsmath,amssymb,amsfonts}
\usepackage{latexsym}
\usepackage{color}
\usepackage{subcaption}
\usepackage{ulem}
\usepackage{mathrsfs}
\usepackage{braket}
\usepackage[table]{xcolor}
\usepackage[usestackEOL]{stackengine} 
 
\usepackage{multirow} 
\usepackage{tensor}
\usepackage{caption}
\usepackage[font=small,labelfont=bf,
   justification=justified,
   format=plain]{caption}

\usepackage{hyperref}
\hypersetup{%
setpagesize=false,
 bookmarksnumbered=true,%
 bookmarksopen=true,%
 colorlinks=true,%
 linkcolor=blue,
 citecolor=blue,
}

\renewcommand{\thesection}{\arabic{section}}

\usepackage{physics}
\usepackage{titlesec}

\titleformat{\section}
  {\normalfont\large\bfseries}{\thesection}{1em}{}
\titlespacing*{\section}
  {0pt}{1.4ex plus .1ex minus .1ex}{1.ex plus .2ex minus .2ex}

\titleformat{\subsection}
  {\normalfont\bfseries}{\thesection}{1em}{}
\titlespacing*{\subsection}
  {0pt}{1.4ex plus .1ex minus .1ex}{1.ex plus .2ex minus .2ex}

\def\nn{\nonumber}

\def\H{\mathcal H}
\def\Rm{\mathcal R_m}
\def\Re{\mathcal R_e}

\begin{document}
\allowdisplaybreaks
\title{\boldmath Baryogenesis in $R^2$-Higgs Inflation: the Gravitational Connection}

\author{Yann Cado}
\affiliation{Laboratoire de Physique Th\'eorique et Hautes Energies (LPTHE), Sorbonne Universit\'e et CNRS UMR 7589, 4 place Jussieu, 75252 Paris CEDEX 05, France\\[0.2cm]}
\author{Christoph Englert}
\affiliation{School of Physics \& Astronomy, University of Glasgow, Glasgow G12 8QQ, UK\\[0.2cm]}
\author{Tanmoy Modak}
\affiliation{Institut f\"ur Theoretische Physik, Universit\"at Heidelberg, 69120 Heidelberg, Germany\\[0.2cm]}
\author{Mariano Quir\'{o}s}
\affiliation{Institut de F\'{i}sica d'Altes Energies (IFAE) and The Barcelona Institute of Science and Technology (BIST),
Campus UAB, 08193 Bellaterra, Barcelona, Spain\\[0.2cm]}

\begin{abstract}
\noindent $R^2$-Higgs inflation stands out as one of the best-fit models of Planck data. Using a covariant formalism for the inflationary dynamics and the production of helical gauge fields, we show that the observed baryon asymmetry of the Universe (BAU) can be obtained when this model is supplemented by a dimension-six CP-violating term $\sim (R/\Lambda^2)\, B_{\mu\nu} \widetilde{B}^{\mu\nu}$  in the hypercharge sector. At linear order, values of $\Lambda\simeq 2.5\times10^{-5}\ M_{\rm P}$ produce, in the $R^2$-like regime, sufficient helical hypermagnetic fields to create the observed matter-antimatter asymmetry during the electroweak crossover.  However, the Schwinger effect of fermion pair production can play a critical role in this context, and that scale is significantly lowered when the backreaction of the fermion fields on the gauge field production is included. In all cases, the helical field configurations can remain robust against washout after the end of inflation.
\end{abstract}
\maketitle
\hrule
\hspace{1ex}
\tableofcontents
\hspace{1ex}
\numberwithin{equation}{section}
\hrule
\setlength{\parskip}{0.5\baselineskip}
\setlength{\parindent}{0pt}
\hspace{1ex}
\section{Introduction}
Cosmic inflation~\cite{Starobinsky:1980te,Sato:1980yn,Guth:1980zm}
elegantly addresses a plethora of observations, ranging from the
flatness of the Universe, over resolving the horizon and exotic relics
problems, all the way to seeding the primordial density perturbations
giving rise to the large-scale structure of the Universe that we see
today. In parallel, it can explain the cosmic microwave background
(CMB) anisotropies measured by experiments such as
Planck~\cite{Planck:2018jri}. While there are several alternatives to
inflation, among these models, Starobinsky or
$R^2$~\cite{Starobinsky:1980te,
  Starobinsky:1983zz,Vilenkin:1985md,Mijic:1986iv, Maeda:1987xf}
inflation, where pure General Relativity (GR) is extended by an
additional scalar curvature term $R^2$, is one of the best-fitting
models of current data~\cite{Planck:2018jri}.

In the dual scalar-tensor theory, the presence of the $R^2$ term makes
the scalar degree of freedom dynamical, which can account for cosmic
inflation. After the discovery of the Higgs boson at the Large Hadron
Collider (LHC)~\cite{ATLAS:2012yve,CMS:2012qbp}, the theory
essentially contains two scalar degrees of freedom. Indeed, if the
Higgs field $\Phi$ couples non-minimally to the Ricci scalar $R$ via a
term $\xi_H R |\Phi|^2$, with $\xi_H$ as the nonminimal coupling, the
Higgs field itself can induce
inflation~\cite{Bezrukov:2007ep,Barvinsky:2008ia,Bezrukov:2010jz,Bezrukov:2013fka,DeSimone:2008ei,Bezrukov:2008ej,Barvinsky:2009ii}
(for earlier works which employed similar mechanisms,
see~\cite{Spokoiny:1984bd,Futamase:1987ua,Salopek:1988qh,Fakir:1990eg,Amendola:1990nn,Kaiser:1994vs,Cervantes-Cota:1995ehs,Komatsu:1999mt}). In
pure Higgs inflation, i.e.~without the presence of such $R^2$ term, a
scale of unitarity violation
emerges~\cite{Burgess:2009ea,Barbon:2009ya,Burgess:2010zq,Hertzberg:2010dc}. This
may not pose a threat to inflationary dynamics, see Ref.~\cite{Antoniadis:2021axu}. However, during the
preheating stage, longitudinal gauge bosons with momenta beyond the
unitarity cut-off scale are
violently produced~\cite{DeCross:2015uza,Ema:2016dny,Sfakianakis:2018lzf}. The
perturbative unitarity is restored up to the Planck scale due to the
presence of $R^2$ term in $R^2$-Higgs inflation~\cite{Ema:2017rqn}
(see also
e.g.~\cite{Salvio:2015kka,Pi:2017gih,Gorbunov:2018llf,Gundhi:2018wyz,He:2018gyf,He:2018mgb,Cheong:2019vzl,Bezrukov:2019ylq,He:2020ivk,Bezrukov:2020txg,He:2020qcb,Aoki:2022dzd}). 
Moreover, $R^2$-Higgs inflation (or the Starobinsky-Higgs inflation), which
features both the $R^2$ and $R |\Phi|^2$ terms, is also the best-fit
model for the Planck data.

Following on from these successes, it is not unreasonable to correlate
the $R^2$-Higgs inflation to the other shortcomings of the current
microscopic theory of interactions, the Standard Model of Particle
Physics (SM). One such shortfall is the observed matter-antimatter
asymmetry (or the Baryon asymmetry) of the Universe, BAU. The
existence of the BAU is a strong indicator of the presence of interactions
beyond the SM.  A range of particle physics experiments, chiefly at
the LHC, are searching for such interactions at the currently largest
available energy scales of ${\cal{O}}(\text{TeV})$. If the fundamental
scale of the mechanism behind the BAU is tied to a higher scale, it might
be possible that tell-tale effects at present or even future colliders
could remain absent. In the SM, the CP-violation from the CKM matrix
is not sufficient for
baryogenesis~\cite{Farrar:1993sp,Farrar:1993hn,Gavela:1993ts}. Further,
the electroweak phase transition in the SM is a continuous
crossover~\cite{Kajantie:1996qd} rather than the typically desired
strong first-order transition to drive the departure from thermal
equilibrium condition as part of Sakharov's
criteria~\cite{Sakharov:1967dj}. However, even at the crossover, the
out-of-equilibrium condition can be met if the source and washout
decay rates are different and shut off at different epochs~\cite{Shaposhnikov:1987tw,Kamada:2016eeb}.
If the inflaton field couples to the CP-odd hypercharge Chern-Simons
density $F\tilde F$, with $F$ and $\tilde F$ denoting the field-stress
tensor of a $U(1)$ gauge field (which mixes with the hypercharge gauge field) and its dual, respectively, helical
hypermagnetic fields can be abundantly produced at the end of
inflation~\cite{Anber:2006xt,Bamba:2006km,Bamba:2007hf,Anber:2009ua,Anber:2015yca,Cado:2016kdp}. The helical
hypermagnetic fields may then create the observed baryon asymmetry at
the electroweak
crossover~\cite{Kamada:2016eeb,Kamada:2016cnb,Jimenez:2017cdr,Domcke:2019mnd,Cado:2021bia,Cado:2022evn,Cado:2022pxk}.

In this article, we investigate baryogenesis in $R^2$-Higgs inflation from CP-violating dimension-six Chern-Simons density
$\sim {(R/ \Lambda^2)} B_{\mu\nu} \widetilde{B}^{\mu\nu}$, where
$R$ is the Ricci scalar and $B_{\mu\nu}$ is the field stress tensor of $U(1)_Y$ hypercharge in the Jordan frame (see also Refs.~\cite{Durrer:2022emo,Durrer:2023rhc,Savchenko:2018pdr,Subramanian:2015lua,Durrer:2013pga} for similar discussions). This term can be considered within the context extended theories of gravity (or
rather, $f(R,\phi, B_\mu)$ gravity), and it elegantly connects high-scale BAU
to inflationary dynamics without requiring additional fields beyond
the SM. Adopting the covariant formalism due to the non-canonical kinetic
terms in $R^2$-Higgs inflation, our linear order analysis, with
$\Lambda\sim 10^{-5}M_{\rm P}$, demonstrates that the produced helical
hypermagnetic fields are sufficient to account for the BAU.
We take into account effects that could lead to a washout of the
helicity stored in the gauge sector (e.g.~the chiral plasma
instability) alongside observational bounds on a range of associated
phenomena that prevent total freedom of the possible field configurations.

In the presence of strong gauge fields, light fermions charged under
the gauge group are produced by the backreaction of gauge fields that
source the fermions equation of
motion~\cite{Domcke:2018eki,Kitamoto:2021wzl}. The corresponding
currents can then, in turn, backreact on the produced gauge fields, a
phenomenon called the \textit{Schwinger effect}, see
e.g.~Ref.~\cite{Cohen:2008wz}.  The backreaction of fermion currents
on the produced gauge fields acts as a damping force during the explosive
production of helical gauge fields, and many of the conclusions from
the gauge field production should be revised in the presence of the
Schwinger effect.  In particular, it has been shown that, although the
amount of gauge energy density is suppressed, which jeopardizes the
gauge preheating capabilities, there is still a window for the
baryogenesis mechanism, see Ref.~\cite{Cado:2022pxk}. Also, one
possible way out is if there are no light, charged fermion fields when gauge
fields are produced, for instance by the use of a special Froggatt-Nielsen
mechanism such that all fermion Yukawa couplings stay large at the end of 
inflation, while they relax after inflation to the measured values~\cite{Cado:2023gan}. 
However, in this paper, we will stay agnostic on the fermions effect in the plasma 
and provide the results with and without the Schwinger effect.

We organize this paper as follows. We start with outlining the action
and derive the relevant equations of motion (EoM) for different fields
in Sec.~\ref{sec:Action}, followed by the inflationary dynamics in the
covariant formalism in Sec.~\ref{sec:infdynamics}. The production of hypermagnetic
fields and subsequent generation of the BAU are discussed, respectively, in
Sec.~\ref{sec:magneto} and Sec.~\ref{sec:baryogensis}. We summarize
with some discussion in Sec.~\ref{disc}. Finally, we present some technical computational 
details through appendices~\ref{vier}-\ref{rk4method}.

\section{The Starobinsky-Higgs Action}\label{sec:Action}
In pure GR with a canonically coupled scalar theory, without the
presence of $R^2$, the conformal mode of the metric is known to have a
wrong-sign kinetic term. The Starobinsky inflation model, which
extends pure GR with an additional scalar curvature term $R^2$, falls
within the so-called general $f(R)$ theory of gravity. In its dual
scalar-tensor theory, the presence of the $R^2$ term makes the scalar
degree of freedom dynamical, which can then account for cosmic
inflation. $R^2$-Higgs inflation (or Starobinsky-Higgs inflation),
which features all possible dimension-four terms i.e.~both the $R^2$
and $R |\Phi|^2$ terms also provide best-fit models of the Planck
data. The model has two dynamical scalar degrees of freedom, one
appearing from the gravity sector and one entering as part of the
Higgs field $\Phi$. 

We briefly discuss the action and its transformation properties in the metric formalism assuming the affine
connection to be the Levi-Civita connection. The action in the Jordan frame of $R^2$-Higgs
inflation, along with a dimension-six CP-odd term coupling Ricci scalar
and $U(1)_Y$ gauge boson, is given by
\begin{equation} \begin{aligned}
  S_J  &=  \int d^4 x \sqrt{-g_J} \bigg[ \frac{M_{\rm P}^{2}}{2} f(R_J, \Phi, B_\mu) 
  -g_J^{\mu\nu}(\nabla_\mu\Phi)^\dagger \nabla_\nu\Phi - V(\Phi, \Phi^\dagger)  -  \dfrac{1}{4} g_J^{\mu\rho} g_J^{\nu\sigma} B_{\mu\nu}B_{\rho\sigma} \\
  & \quad \quad \quad \quad \quad \quad \quad - \dfrac{1}{4} g_J^{\mu\rho} g_J^{\nu\sigma} W^i_{\mu\nu}W^i_{\rho\sigma}
  - \sum_f g_J^{\mu\nu} \bar f e^a_{\mu} \tilde{\gamma}_a \nabla^f_{\nu}  f \bigg],\label{eq:actionJ1}
\end{aligned} \end{equation}
and we adopt a mostly-plus convention for the metric $(-1,+1,+1,+1)$. From here on, for notational simplicity, we will remove the sum over fermions $\sum_f$ in the fermion quadratic terms, which will remain implicit. The $B_{\mu\nu}$ and $W^i_{\mu\nu}$ are field stress tensors of the $U(1)_Y$ and $SU(2)_L$ gauge groups, respectively, $\Phi$ is the Higgs field, $R_J$ the Ricci scalar in the Jordan frame, and $M_{\rm P}=\sqrt{1/\left(8\pi G\right)}=2.435\times 10^{18}$ GeV, where $G$ is the Newton's constant and $M_{\rm P}$ the reduced Planck mass. We use the convention $\epsilon^{0123}=1$ for the Levi-Civita tensor. The covariant derivatives are defined as
\begin{subequations} \begin{eqnarray}
\nabla_\mu &=& D_\mu + i g' \frac{1}{2} Q_{Y_f} B_\mu + i g \frac{1}{2}\bm{\tau} \cdot \bm{W}_\mu  ,\\
\nabla^f_{\mu} f &=& \left(D_\mu^f + i g' \frac{1}{2} Q_{Y_f} B_\mu + i g \frac{1}{2} \bm{\tau} \cdot \bm{W}_\mu\right) f,
\end{eqnarray} \end{subequations}
with $Q_{Y_f}$ denoting the $U(1)_Y$ hypercharge, $\bm{\tau}$ are the Pauli matrices, and $g'$ and $g$ are
respective gauge couplings. $D_\mu$ is the usual covariant
derivative with respect to the space-time metric $g_{J\mu\nu}$ and $D_\mu^f \equiv \partial_\mu + \Gamma_\mu $ is the covariant derivative of spinors, with 
$\Gamma_\mu$ as the spin affine connection. Here $e^a_{\mu}$ is the so-called vierbein and $\tilde{\gamma}_a$ is Minkowski space gamma matrices (see Appendix~\ref{vier} for details 
of the formalism and the definition of $\Gamma_\mu$). 
The corresponding field-stress tensors for the $U(1)_Y$ and $SU(2)_L$ gauge fields are
\begin{align}
B_{\mu\nu} = D_\mu B_\nu - D_\nu B_\mu, \quad \quad W^i_{\mu\nu} = D_\mu W^i_\nu - D_\nu W^i_\mu - g \sum_{j,k=1}^{3} \epsilon_{ijk} W^j_\mu W^k_\nu.
\end{align}
The Higgs potential $V(\Phi, \Phi^\dagger)$ and $f(R_J, \Phi, B_\mu)$ are given as\footnote{For large configuration values of the Higgs field we can consistently neglect the mass term of the Higgs potential, which triggers electroweak symmetry breaking.}
\begin{subequations} \begin{eqnarray}
V(\Phi, \Phi^\dagger) &=& \lambda|\Phi|^4, \\ f(R_J, \Phi, B_\mu) &=& R_J  + \frac{\xi_R}{2 M_{\rm P}^2} R_J^2 + \frac{2\xi_H}{M_{\rm P}^2} |\Phi|^2 R_J
- \frac{2}{\Lambda^2 M_{\rm P}^2}\frac{\epsilon^{\mu\nu\rho\sigma}}{\sqrt{-g_J}} B_{\mu\nu} B_{\rho\sigma} R_J.
\end{eqnarray} \end{subequations}
The Higgs field has hypercharge $+1$ and is decomposed in the standard
way (we will comment on our gauge choice further below) 
\begin{align}
\Phi =
\frac{1}{\sqrt{2}} \begin{pmatrix}
  0 \\
  h \\
\end{pmatrix} \label{eq:higgsuni}.
\end{align}
With this choice, Eq.~\eqref{eq:actionJ1} becomes
\begin{equation} \begin{aligned}
  S_J  = &  \int d^4 x \sqrt{-g_J} \bigg[\frac{M_{\rm P}^{2}}{2} f(R_J, h, B_\mu) 
  -\frac{1}{2} g_J^{\mu\nu} (D_\mu h) D_\nu h - V(h) \\
 & - \frac{1}{4} g_J^{\mu\nu} g^2 h^2 \frac{(W^1_\mu-i W^2_\mu)}{\sqrt{2}} \frac{(W^1_\nu+i W^2_\nu)}{\sqrt{2}}
  - \frac{1}{8} g_J^{\mu\nu} h^2 \left( g W^3_\mu - g' B_\mu \right) \left( g W^3_\nu - g' B_\nu \right)
  \\   
 & -  \dfrac{1}{4} g_J^{\mu\rho} g_J^{\nu\sigma} B_{\mu\nu}B_{\rho\sigma} 
  - \dfrac{1}{4} g_J^{\mu\rho} g_J^{\nu\sigma} W^i_{\mu\nu}W^i_{\rho\sigma} 
  - g_J^{\mu\nu} \bar f e^a_{\mu} \tilde{\gamma}_a \nabla^f_{\nu}  f\bigg]\label{eq:actionJ2},
\end{aligned} \end{equation}
with
\begin{align}
V(h) = \frac{\lambda}{4} h^4,~~~f(R_J,  h, B_\mu)= R_J  + \frac{\xi_R}{2 M_{\rm P}^2} R_J^2 + \frac{\xi_H}{M_{\rm P}^2}  h^2 R_J 
- \frac{2}{\Lambda^2 M_{\rm P}^2}\frac{\epsilon^{\mu\nu\rho\sigma}}{\sqrt{-g_J}} B_{\mu\nu} B_{\rho\sigma} R_J.
\end{align}
The dynamics of the scalar degrees of freedom are easily captured once
we move from the Jordan frame to the Einstein frame via a Weyl
transformation. We first introduce an auxiliary field $\Psi$ and
rewrite the action in Eq.~\eqref{eq:actionJ2} as
\begin{equation} \begin{aligned}
  S_J  = & \int d^4 x \sqrt{-g_J} \bigg[\frac{M_{\rm P}^{2}}{2} \left(f(\Psi,  h, B_\mu) 
  + \frac{\partial f(\Psi,  h, B_\mu)}{\partial \Psi} (R_J-\Psi)\right)
  -\frac{1}{2} g_J^{\mu\nu}(D_\mu h) (D_\nu h) - V( h)   
  \\
  & - \frac{1}{4} g_J^{\mu\nu} g^2 h^2 \frac{(W^1_\mu-i W^2_\mu)}{\sqrt{2}} \frac{(W^1_\nu+i W^2_\nu)}{\sqrt{2}}
  - \frac{1}{8} g_J^{\mu\nu} h^2 \left( g W^3_\mu - g' B_\mu \right) \left( g W^3_\nu - g' B_\nu \right)
   \\
 & - \dfrac{1}{4} g_J^{\mu\rho} g_J^{\nu\sigma} B_{\mu\nu}B_{\rho\sigma} - \dfrac{1}{4} g_J^{\mu\rho} g_J^{\nu\sigma} W^i_{\mu\nu}W^i_{\rho\sigma}
  -  g_J^{\mu\nu} \bar f e^a_{\mu} \tilde{\gamma}_a \nabla^f_{\nu}  f\bigg]\label{eq:actionJ3}.
\end{aligned} \end{equation}
The variation with respect to $\Psi$ gives the constraint $\Psi = R_J$
as long as $\partial^2 f(\Psi, h, B_\mu)/\partial \Psi^2\neq 0$.
We now define a physical degree of freedom $\Theta$ as
\begin{align}
\Theta = \frac{\partial f(\Psi,  h, B_\mu)}{\partial \Psi}
\end{align}
such that the action Eq.~\eqref{eq:actionJ3} can be cast into
\begin{equation} \begin{aligned}
  S_J  =&  \int d^4 x \sqrt{-g_J} \bigg[\frac{M_{\rm P}^{2}}{2} \Theta R_J - U(\Theta,   h, B_\mu)
  - \frac{1}{2} g_J^{\mu\nu}(D_\mu h) (D_\nu h) - V( h)  -  \dfrac{1}{4} g_J^{\mu\rho} g_J^{\nu\sigma} B_{\mu\nu}B_{\rho\sigma}\\
& - \frac{1}{4} g_J^{\mu\nu} g^2 h^2 \frac{(W^1_\mu-i W^2_\mu)}{\sqrt{2}} \frac{(W^1_\nu+i W^2_\nu)}{\sqrt{2}}
  - \frac{1}{8} g_J^{\mu\nu} h^2 \left( g W^3_\mu - g' B_\mu \right) \left( g W^3_\nu - g' B_\nu \right)\\
 & - \dfrac{1}{4} g_J^{\mu\rho} g_J^{\nu\sigma} W^i_{\mu\nu}W^i_{\rho\sigma}
  - g_J^{\mu\nu} \bar f e^a_{\mu} \tilde{\gamma}_a \nabla^f_{\nu}  f\bigg]\label{eq:actionJ4},
\end{aligned} \end{equation}
with the definition
\begin{equation} \begin{aligned}
U(\Theta,   h, B_\mu) &= \frac{M_{\rm P}^2}{2}\left[\Psi(\Theta) \Theta -  f(\Psi(\Theta),  h, B_\mu)\right] \\ &= \frac{M_{\rm P}^4}{4 \xi_R}\left(1- \Theta + \frac{\xi_H}{M_{\rm P}^2} h^2 
                          - \frac{2}{\Lambda^2 M_{\rm P}^2}\frac{\epsilon^{\mu\nu\rho\sigma}}{\sqrt{-g_J}} B_{\mu\nu} B_{\rho\sigma}\right)^2.
\end{aligned} \end{equation}
To formulate the action in the Einstein frame, we perform the metric redefinition (Weyl transformation)
\begin{align}
g_{J\mu\nu} = \frac{1}{\Theta} \ g_{E\mu\nu},~~~g^{\mu\nu}_J = \Theta \ g^{\mu\nu}_E, ~\mbox{and}~\sqrt{-g_{J}}= \frac{1}{\Theta^2}\sqrt{-g_E}.
\end{align}
Under this transformation, the Ricci scalar transforms as
\begin{align}
R_J = \Theta \left[R_E +3 \Box_E  \Theta- \frac{3}{2} g_{E}^{\mu\nu} D_\mu (\ln\Theta) D_\nu(\ln\Theta) \right],
\end{align}
with $\Box_E = g_{E}^{\mu\nu} D_\mu D_\nu$.  Ignoring the surface
term, the action of Eq.~\eqref{eq:actionJ3} now becomes
\begin{equation} \begin{aligned}
S_E  =& \int d^4 x \sqrt{-g_E}\bigg[\frac{M_{\rm P}^2}{2} R_E - \frac{3 M_{\rm P}^2}{4} g_E^{\mu \nu} D_\mu (\ln\Theta) D_\nu(\ln\Theta)
-\frac{1}{2 \Theta} g_E^{\mu\nu}(D_\mu h) (D_\nu h) - V_E \\
& -  \dfrac{1}{4} g_E^{\mu\rho} g_E^{\nu\sigma} B_{\mu\nu}B_{\rho\sigma}   - \frac{1}{4 \Theta} g_E^{\mu\nu} g^2 h^2 \frac{(W^1_\mu-i W^2_\mu)}{\sqrt{2}} \frac{(W^1_\nu+i W^2_\nu)}{\sqrt{2}}
\\
 & - \frac{1}{8 \Theta} g_E^{\mu\nu} h^2 \left( g W^3_\mu - g' B_\mu \right) \left( g W^3_\nu - g' B_\nu \right)- \dfrac{1}{4} g_E^{\mu\rho} g_E^{\nu\sigma} W^i_{\mu\nu}W^i_{\rho\sigma}
  - \frac{1}{\Theta} g_E^{\mu\nu}\bar f e^a_{\mu} \tilde{\gamma}_a \nabla^f_{\nu}  f\bigg], 
\end{aligned} \end{equation}
with 
\begin{align}
V_E = \frac{1}{\Theta^2}\left[V( h)+U(\Theta,   h, B_\mu)\right].
\end{align}
Finally, we perform the field redefinition 
\begin{align}
\phi = M_{\rm P} \sqrt{\frac{3}{2}} \ln\Theta.
\end{align}
to arrive at the action in the form
\begin{equation} \begin{aligned}
S_E  =& \int d^4 x \sqrt{-g_E}\;\bigg[\frac{M_{\rm P}^2}{2} R_E - \frac{1}{2} G_{IJ} g_E^{\mu \nu} D_\mu \phi^I D_\nu \phi^J - 
V_E(\phi^I) -  \dfrac{1}{4} g_E^{\mu\rho} g_E^{\nu\sigma} B_{\mu\nu}B_{\rho\sigma}  \\
 &- \dfrac{1}{4} g_E^{\mu\rho} g_E^{\nu\sigma} W^i_{\mu\nu}W^i_{\rho\sigma} - \frac{1}{4 } e^{-\sqrt{\frac{2}{3}}\frac{\phi}{M_{\rm P}}} g_E^{\mu\nu} g^2 h^2 \frac{(W^1_\mu-i W^2_\mu)}{\sqrt{2}} \frac{(W^1_\nu+i W^2_\nu)}{\sqrt{2}}\\
  &-  \frac{1}{8 } e^{-\sqrt{\frac{2}{3}}\frac{\phi}{M_{\rm P}}} g_E^{\mu\nu} h^2 \left(g W^3_\mu - g' B_\mu \right) \left( g W^3_\nu - g' B_\nu \right)
  -  e^{-\sqrt{\frac{2}{3}}\frac{\phi}{M_{\rm P}}} g_E^{\mu\nu}\bar f e^a_{\mu} \tilde{\gamma}_a \nabla^f_{\nu}  f \bigg]. \label{eq:actionfinal}
\end{aligned} \end{equation}
The multi-field $\phi^I \in \{\phi, h\}$ alongside the field-space metric $G_{IJ}$ 
\begin{align}
G_{\phi\phi} = 1,~ G_{\phi h} = G_{ h\phi} = 0, ~G_{ h h} = e^{-\sqrt{\frac{2}{3}}\frac{\phi}{M_{\rm P}}}
\end{align}
highlight that we are working with a non-canonical kinetic term as
alluded to above (see Appendix~\ref{fieldchris} for the corresponding field-space Christoffel symbols). The potential $V_E(\phi^I)$, consistently truncated at dimension-six level, reads
\begin{align}
V_E(\phi^I) =& \;e^{-2\sqrt{\frac{2}{3}}\frac{\phi}{M_{\rm P}}} \bigg[\frac{\lambda}{4} h^4 + \frac{M_{\rm P}^4}{4 \xi_R}\bigg(1- e^{\sqrt{\frac{2}{3}}\frac{\phi}{M_{\rm P}}}  + \frac{\xi_H}{M_{\rm P}^2} h^2 
                          - \frac{2}{\Lambda^2 M_{\rm P}^2}\frac{\epsilon^{\mu\nu\rho\sigma}}{\sqrt{-g_E}} e^{2 \sqrt{\frac{2}{3}}\frac{\phi}{M_{\rm P}}}  B_{\mu\nu} B_{\rho\sigma}\bigg)^2\bigg]\nn\\                             
=& \;V_0(\phi^I)+\frac{2 M_{\rm P}^2}{\xi_R\Lambda^2} 
F(\phi^I)e^{\sqrt{\frac{2}{3}}\, \frac{\phi}{M_{\rm P}}}B_{\mu\nu}\tilde B^{\mu\nu},
\end{align}
with
\begin{subequations} 
\begin{eqnarray}
V_0(\phi^I)&=& \frac{\lambda}{4}h^4e^{-2\sqrt{\frac{2}{3}}\,\frac{\phi}{M_{\rm P}}}+\frac{M_{\rm P}^4}{4\xi_R}F^2(\phi^I),\\
F(\phi^I)&=&1-e^{-\sqrt{\frac{2}{3}}\, \frac{\phi}{M_{\rm P}}}+\frac{\xi_H}{M_{\rm P}^2}h^2e^{-\sqrt{\frac{2}{3}}\, \frac{\phi}{M_{\rm P}}},~~\tilde B^{\mu\nu}=\frac{1}{2 \sqrt{-g_E}}\epsilon^{\mu\nu\rho\sigma}B_{\rho\sigma}.
\end{eqnarray} 
\end{subequations}
Note that the unmodified Starobinsky potential is recovered when the term that contains $B_{\mu\nu}\tilde B^{\mu\nu}$ is absent.

We can now turn to the EoMs of the different fields in
Eq.~\eqref{eq:actionfinal}. By varying Eq.~\eqref{eq:actionfinal} with
respect to the field $\phi$, we obtain
\begin{align}
&\Box \phi^K + \Gamma^{K}_{IJ} \ g_E^{\alpha \nu} D_\alpha  \phi^I D_\nu \phi^J - G^{KM} V_{E,M} + g_E^{\mu\nu}\mathcal{X}^K_{\mu\nu}= 0\label{eom:scalar},
\end{align}
identifying $\Gamma^{K}_{IJ}$ as the field-space Christoffel symbols and
\begin{equation} \begin{aligned}
\mathcal{X}^K_{\mu\nu} &=\frac{1}{4 } \sqrt{\frac{2}{3}}\frac{1}{M_{\rm P}} G^{K1} 
e^{-\sqrt{\frac{2}{3}}\frac{\phi}{M_{\rm P}}}  g^2 h^2 \frac{(W^1_\mu-i W^2_\mu)}{\sqrt{2}} \frac{(W^1_\nu+i W^2_\nu)}{\sqrt{2}}\\
&-\frac{1}{2} e^{-\sqrt{\frac{2}{3}}\frac{\phi}{M_{\rm P}}} G^{K2}  g^2 h \frac{(W^1_\mu-i W^2_\mu)}{\sqrt{2}} \frac{(W^1_\nu+i W^2_\nu)}{\sqrt{2}}\\
&+ \frac{1}{8} \sqrt{\frac{2}{3}}\frac{1}{M_{\rm P}} G^{K1} e^{-\sqrt{\frac{2}{3}}\frac{\phi}{M_{\rm P}}}  h^2 \left( g W^3_\mu - g' B_\mu \right) \left( g W^3_\nu - g' B_\nu \right)\\
&- \frac{1}{4} G^{K2} e^{-\sqrt{\frac{2}{3}}\frac{\phi}{M_{\rm P}}}  h \left( g W^3_\mu - g' B_\mu \right) \left( g W^3_\nu - g' B_\nu \right)
 \\
&  + \sqrt{\frac{2}{3}}\frac{1}{M_{\rm P}}  \delta^{~K}_{1} e^{-\sqrt{\frac{2}{3}}\frac{\phi}{M_{\rm P}}} g_E^{\mu\nu} \bar f t^a_\mu\tilde{\gamma}_a \nabla^f_\nu  f.
\end{aligned} \end{equation}
Note that all the terms in $\mathcal{X}^K_{\mu\nu}$ are quadratic
in the gauge fields.

The energy-momentum tensor $T_{\mu\nu}$ describes relevant quantities
of the inflationary dynamics such as energy density or pressure.  One
can derive the Einstein-Hilbert equation from the action $S_E$ by varying it
with respect to $g_E^{\mu\nu}$
\begin{align}
R_{E\mu\nu} -\frac{1}{2} g_{E\mu\nu} R_E &= \frac{1}{M_{\rm P}^2} \left(\mathcal{L}_M g_{E\mu\nu} - 2 \frac{\delta (\mathcal{L}_M)}{\delta g_E^{\mu\nu}}\right)
\end{align}
and identify $T_{\mu\nu}$ as
\begin{align}
T_{\mu\nu} = \left(\mathcal{L}_M g_{E\mu\nu} - 2 \frac{\delta (\mathcal{L}_M)}{\delta g_E^{\mu\nu}}\right).\label{eq:tmunu}
\end{align}
Appendix~\ref{energymomtensor} provides the full expression of
$T_{\mu\nu}$ for the model considered in this work.

The EoM for the gauge field $B_\mu$ is given as 
\begin{equation} \begin{aligned}
&g_E^{\mu \alpha} g_E^{\nu \beta} D_\alpha B_{\mu\nu} +\frac{8 M_{\rm P}^2}{\xi_R \Lambda^2} D_\alpha\left( F(\phi^I)e^{\sqrt{\frac{2}{3}}\, \frac{\phi}{M_{\rm P}}} \right) \tilde{B}^{\alpha\beta}
+\left[\frac{g'}{4} e^{-\sqrt{\frac{2}{3}}\frac{\phi}{M_{\rm P}}} g_E^{\mu\beta} h^2 (g W_\mu^3-g' B_\mu) 
\right]\\  
&\qquad\qquad\qquad\qquad\qquad\qquad\qquad\qquad\qquad\qquad\qquad- \frac{ig' Q_{Y_f}}{2} e^{-\sqrt{\frac{2}{3}}\frac{\phi}{M_{\rm P}}} g_E^{\mu\beta} \bar f e^a_\mu\tilde{\gamma}_a f =0\label{eq:Bmu},
\end{aligned}\end{equation} 
and those for the $W_\mu^i$ fields are found to be
\begin{multline}
g_E^{\mu\alpha} g_E^{\nu\beta} D_\alpha W_{\mu\nu}^i - g g_E^{\mu\beta} g_E^{\nu\sigma} \sum_{j,k=1}^3\epsilon_{i j k} 
W_{\mu\nu}^j  W_\sigma^k-\frac{g}{4} g_E^{\mu\beta} e^{-\sqrt{\frac{2}{3}}\frac{\phi}{M_{\rm P}}} h^2 \mathcal{W}_\mu^i +\frac{ig}{2} e^{-\sqrt{\frac{2}{3}}\frac{\phi}{M_{\rm P}}} g_E^{\mu\beta} \bar f e^a_\mu\tilde{\gamma}_a  \tau^i  f = 0,\\\mbox{with}~ 
\mathcal{W}_\mu^i=
\begin{cases}
    g W_\mu^i, \text{if }~i= 1,2\\
    g W_\mu^3 -g' B_\mu,  \text{if}~i= 3 
\end{cases}.
\label{eq:Wmu}
\end{multline}
We define $W_\mu$, $Z_\mu$ and $A_\mu$ in the usual way
\begin{equation} \begin{aligned}
W_\mu &= \frac{W^1_{\mu} - i W^2_{\mu}}{\sqrt{2}},~~ W^\dagger_\mu = \frac{W^1_{\mu} + i W^2_{\mu}}{\sqrt{2}}\\
A_\mu &= \sin\theta_W W^3_\mu + \cos\theta_W B_\mu,~~ Z_\mu = \cos\theta_W W^3_\mu - \sin\theta_W B_\mu,
\end{aligned} \end{equation}
with $e = g \sin\theta_W = g' \cos\theta_W$, and electroweak angle $\theta_W$. We can express the $W^i_\mu$ and $B_\mu$ fields in terms of  $W_\mu$, $Z_\mu$ and $A_\mu$ 
by inverting the above equations.

Given that we are in the broken phase, for which $h_0\neq 0$, where $h_0$ is the homogeneous background field as we shall see shortly, we can consider the trivial solution
$\mathcal W_\mu^i=0$ from the mass term in Eq.~\eqref{eq:Wmu} as the variation is small compared to the background field. This means that we can set $W_\mu=W_\mu^\dagger =0$ and $Z_\mu=\cos\theta_W W_\mu^3-\sin\theta_W B_\mu=0$ which implies that $B_\mu=\cos\theta_W A_\mu$ and $W^3_\mu=\sin\theta_W A_\mu$. We will therefore retain only the photon field $A_\mu$, replacing $B_\mu$ with $\cos\theta_W A_\mu$ in the corresponding Chern-Simons term. Put differently, the production of photon fields proceeds unsuppressed compared to the other heavy gauge bosons.

We now turn to some comments related to the gauge fixing in Eq.~\eqref{eq:higgsuni}. The Higgs doublet contains, apart from the radial degree of freedom $h$, three Goldstone bosons $\vec\chi$. Using $SU(2)_L$ gauge invariance, and fixing the corresponding gauge parameter $\vec\alpha(x)$ as $\vec\alpha(x)=-\vec \chi(x)$ (unitary gauge), the Goldstone bosons disappear from the Lagrangian and the Higgs doublet reduces to Eq.~(\ref{eq:higgsuni}). There is still the $U(1)$ gauge invariance that can be used to fix the Coulomb gauge for the electromagnetic field $\partial^i A_i=0$. This is done by fixing the hypercharge gauge field $B_\mu$ as $\partial^i B_i=-\tan{\theta_W} \partial^i W_i^3$. Moreover, in regions where the electric charge density is zero, it turns out that $A_0=0$ (the radiation gauge we use in this paper). Therefore, the EoM for the $A_\mu$ field simplifies to
\begin{equation}
\begin{aligned}
g_E^{\mu \alpha} g_E^{\nu \beta} D_\alpha F_{A\mu\nu} 
+ \frac{8\cos^2\theta_W M_{\rm P}^2}{\xi_R \Lambda^2 } \partial_\alpha&\left( F(\phi^I)e^{\sqrt{\frac{2}{3}}\, \frac{\phi}{M_{\rm P}}} \right)
  \tilde{F}_{A}^{\alpha\beta} \\ & = ie Q_f \,e^{-\sqrt{\frac{2}{3}}\frac{\phi}{M_{\rm P}}} g_E^{\mu\beta} \bar f e^a_\mu\tilde{\gamma}_a f , \label{eq:Bmu-broken} 
\end{aligned}
\end{equation}
with $Q_f= \frac{1}{2}Q_{Y_f}+T_{3 f}$, where $T_3$ is the third component of weak isospin.

Similarly, one can find the general covariant Dirac equation as 
\begin{align}
g_E^{\mu\nu}   e^a_\mu\tilde{\gamma}_a (\nabla^f_\nu f) =0.
\end{align}

\section{Inflationary Dynamics in the Covariant Formalism}\label{sec:infdynamics}
We now study the inflationary dynamics of our two-field scenario with the non-canonical kinetic term (i.e.~with a nontrivial field-space manifold) following the covariant formalism discussed in Refs.~\cite{Gong:2011uw,Kaiser:2012ak,Sfakianakis:2018lzf} (see also Refs.~\cite{Sasaki:1995aw,Gordon:2000hv,GrootNibbelink:2000vx,GrootNibbelink:2001qt,Wands:2002bn,Seery:2005gb,Peterson:2010np,Peterson:2011yt,Elliston:2012ab,DeCross:2015uza,DeCross:2016fdz,DeCross:2016cbs,Lee:2021rzy}). Focussing on linear order perturbations, we decompose the fields into classical background (${\varphi}^I $) and perturbation parts ($\delta\phi^I$) as
\begin{align}
\label{fieldexpan}
\phi^I(x^\mu) = \varphi^I(t) + \delta\phi^I(x^\mu),
\end{align}
with $\varphi^I(t) = \{\varphi(t),h_0(t)\}$. The space-time dynamics
can be described by the perturbed spatially flat
Friedmann-Robertson-Walker (FRW) metric, which is expanded
as~\cite{Kodama:1984ziu,Mukhanov:1990me,Malik:2008im}
\begin{equation}
ds^2 = -(1+2\mathcal{A}) dt^2 +2 a(t) (\partial_i \mathcal{B}) dx^i dt +
a(t)^2 \left[(1-2\psi) \delta_{ij}+ 2 \partial_i \partial_j \mathcal{E}\right] dx^i dx^j.
\end{equation}
$a(t)$ denotes the scale factor, $t$ parametrizes cosmic time, and
$\mathcal{A}, \mathcal{B}, \psi$ and $\mathcal{E}$ characterize the
scalar metric perturbations. Like the scalar fields, the space-time
metric is also considered up to first order in the perturbations. In the
following, when deriving the background and perturbation equations for
scalar and gauge fields, we shall adopt the longitudinal gauge, i.e.~$\mathcal{B}=\mathcal{E}=0$.

One may define covariant field fluctuations $\mathcal{Q}^I$ (covariant with respect to the field-space metric) that connect $\phi^I(x^\mu)$
and $\varphi^I(t)$ along the geodesic of the field-space manifold with
affine connection~$\kappa$. Concretely, we can take
$\phi^I(\kappa = 0) =\varphi^I$,
$\phi^I(\kappa = \kappa') = \varphi^I +\delta \phi^I$ and
$D_\kappa\phi^I\vert_{\kappa=0}= {d\phi^I}/{d\kappa}\vert_{\kappa=0}
\equiv \mathcal{Q}^I$, such that with these conditions, the unique
field-space vector $\mathcal{Q}^I$ connects $\phi^I$ and
$\varphi^I$~\cite{Gong:2011uw}. Note here, that $D_\kappa$ is the
covariant derivative with respect to the affine connection. The field
fluctuations $\delta\phi^I$ can be expressed in a series of
$\mathcal{Q}^I$ as~\cite{Gong:2011uw,Elliston:2012ab}
\begin{align}
\label{deltaphiI}
\delta\phi^I &= \mathcal{Q}^I -\frac{1}{2} \Gamma^I_{JK} \mathcal{Q}^I \mathcal{Q}^J+\frac{1}{3!} 
\big(\Gamma^I_{MN} \Gamma^N_{JK}-\Gamma^I_{JK,M}\big) \mathcal{Q}^I \mathcal{Q}^J  \mathcal{Q}^M+\dots,
\end{align}
where the Christoffel symbols $\Gamma^I_{JK}$ are evaluated at the
background field order. The field fluctuations $\delta\phi^I(x^\mu)$
are gauge-dependent quantities under both the field-space
transformation $\varphi^I \to \varphi'^{I}$, as well as the space-time
transformation $x^\mu\to x'^{\mu}$. This is motivation to formulate
gauge-independent Mukhanov-Sasaki variables, which are a linear
combinations of space-time metric perturbation $\psi$ and covariant
field fluctuations $\mathcal{Q}^I$
as~\cite{Sasaki:1986hm,Mukhanov:1988jd,Mukhanov:1990me}
\begin{align}
Q^I = \mathcal{Q}^I + \frac{\dot{\varphi}^I}{H}\psi \label{mukh-sasaki}.
\end{align}
We remark that, while $\varphi^I$ is not a vector of the field-space 
manifold, $\mathcal{Q}^I$, $\dot{\varphi}^I$ and $Q^I$ all transform, indeed, 
as vectors of the field-space manifold. The $Q^I$ is doubly covariant with 
respect to both space-time and field-space transformations to first order in the perturbations. It is useful to define the covariant derivative of vectors 
$S^I$ and $S_I$ in the field-space as 
\begin{align}
\mathcal{D}_J S^I \equiv \partial_J S^I + \Gamma^I_{JK} S^K, 
\qquad
\mathcal{D}_J S_I \equiv \partial_J S_I - \Gamma^K_{IJ} S_K.
\end{align}
It is convenient to also define a covariant derivative with respect to cosmic time $t$
\begin{align}
\mathcal{D}_t S^I \equiv \dot{\varphi}^J \mathcal{D}_J S^I = \dot{S}^I + \Gamma^I_{JK} S^J \dot{\varphi}^K,
\end{align}
see also Refs.~\cite{Easther:2005nh,Langlois:2008mn,Peterson:2010np,Peterson:2010mv,Peterson:2011yt}.

We turn to the stress-energy tensor $T_{\mu\nu}$, which can be written
for the homogeneous, isotropic and spatially flat metric
$\tilde{g}_{E\mu\nu}={\rm diag}(-1,a^2(t), a^2(t), a^2(t))$ as
\begin{align}
T_{\mu\nu} = (p+\rho) U_\mu U_\nu + p\, g_{\mu\nu},\label{tmunufluid}
\end{align}
with a choice of $U_\mu =(1,0,0,0)$ for the fluid four-velocity. For a
spatially flat metric, employing Eq.~\eqref{tmunufluid} and the
Einstein equations, we get the Friedmann equations for the background
order
\begin{align}
H^2 &= \left(\frac{\dot{a}}{a}\right)^2 = \frac{1}{3 M_{\rm P}^2} \rho,~~\mbox{and}~~\dot{H} = -\frac{1}{2 M_{\rm P}^2} (p+\rho),
\end{align}
where $p$ and $\rho$ are pressure and energy density, respectively.
We can compare the $00$ and $ij$ component of Eq.~\eqref{eq:tmunu} and
Eq.~\eqref{tmunufluid} to get expressions for pressure $p$ and energy
density $\rho$,
\begin{align}
\rho = T_{00},~~p  = \frac{1}{3 a^2}\sum_{i=1}^3 T_{ii}.
\end{align} 
At the considered background order, employing the explicit expression
of Eq.~\eqref{eq:energymomten} (see Appendix~\ref{energymomtensor}), the (inflaton)
pressure and energy density reduce to
\begin{subequations} \begin{eqnarray}
\rho
&=& \frac{1}{2} G_{IJ} \dot{\varphi}^I \dot{\varphi}^J + V_0(\varphi^I),\label{eq:infenergy}
\\
p &=& \frac{1}{2} G_{IJ} \dot{\varphi}^I \dot{\varphi}^J - V_0(\varphi^I),
\end{eqnarray} \end{subequations}
yielding the equation of state
\begin{align}
w =\frac{p}{\rho} =\frac{ G_{IJ} \dot{\varphi}^I \dot{\varphi}^J - 2 V_0}{G_{IJ} \dot{\varphi}^I \dot{\varphi}^J + 2 V_0}.
\end{align}
Furthermore, the Hubble parameter and its derivative with respect to cosmic time take the form
\begin{subequations} \begin{eqnarray}
H^2 &=& \left(\frac{\dot{a}}{a}\right)^2 = \frac{1}{3 M_{\rm P}^2} \left(\frac{1}{2} G_{IJ} \dot{\varphi}^I \dot{\varphi}^J + V_0(\varphi^I)\right),\label{hubble1}\\
\dot{H} &=& -\frac{1}{2 M_{\rm P}^2} \left(G_{IJ} \dot{\varphi}^I \dot{\varphi}^J\right)\label{hubble2}.
\end{eqnarray} \end{subequations}
The EoMs for the background fields $\varphi^I$ and the perturbations
$Q^I$ at linear order can be derived utilizing
Eq.~\eqref{mukh-sasaki}, and Eq.~\eqref{eom:scalar} 
\begin{subequations} \begin{eqnarray}
&&\mathcal{D}_t \dot{\varphi}^I + 3 H\dot{\varphi}^I + G^{IJ} V_{0,J}= 0\label{eq:bkg},\\
&&\mathcal{D}_t^2 Q^I + 3 H D_t Q^I +\frac{k^2}{a^2} \delta^I_J Q^J + \mathcal{M}^I_L Q^L=0,\label{eq:fluc}
\end{eqnarray} \label{eq:bkg+fluc}\end{subequations}
with
\begin{align}
\mathcal{M}^{I}_{L} = G^{IJ} (\mathcal{D}_L\mathcal{D}_J V_0)- \mathcal{R}^I_{JKL} \dot{\varphi}^J \dot{\varphi}^K
- \frac{1}{M_{\rm P}^2 a^3} \mathcal{D}_t \left(\frac{a^3}{H}\dot{\varphi}^I \dot{\varphi}_L\right),\label{eq:massterm}
\end{align}
and the field-space Riemann tensor $\mathcal{R}^I_{JKL}$. All relevant
quantities such as $V_0$, $G^{IJ}$, $\Gamma^{I}_{JK}$,
$\mathcal{R}^I_{JKL}$ in Eqs.~\eqref{eq:bkg+fluc} are evaluated at background order.  Moreover, as the field-space metric $G^{IJ}$ and $M^{IJ}$ are diagonal in this approximation, the first-order perturbations do not mix the different $Q^I$.  Note also that the EoMs for background and perturbations do not depend on the gauge fields for our linear-order considerations.

To study perturbations, we can find a set of unit vectors that differentiate between adiabatic and entropy directions. Firstly, we define the length of the velocity vector $\dot{\varphi}^I$ in field-space defined as 
\begin{align}
\dot{\sigma} = \sqrt{G_{IJ} \dot{\varphi}^I \dot{\varphi}^J} = \sqrt{\rho+p}
\label{eq:sigmadot}
\end{align}
and the corresponding unit vector
\begin{align}
 \hat{\sigma}^I = \frac{\dot{\varphi}^I}{\dot\sigma}.
\label{eq:sigmadotI}
\end{align}
With this, we can rewrite Eq.~\eqref{eq:bkg} to reproduce
a single-field model with a canonically normalized kinetic
term. The slow-roll parameters $\epsilon$ and $\eta$
are
\begin{subequations} \begin{eqnarray}
\epsilon &=& -\frac{\dot{H}}{H^2}=\frac{3\dot{\sigma}^2}{\dot{\sigma}^2+2V_0},\label{epsi}\\
\eta &=& M_{\rm P}^2 \frac{\mathcal{M}_{\sigma\sigma}}{V_0}, 
\end{eqnarray} \end{subequations}
with
$\mathcal{M}_{\sigma\sigma} \equiv \hat{\sigma}_I \hat{\sigma}^J
\mathcal{M}^I_J = \hat{\sigma}^I \hat{\sigma}^J
(\mathcal{D}_I\mathcal{D}_J V_0)$. Inflation ends when the slow-roll
parameter reaches $\epsilon=1$, and we denote the corresponding
cosmological time as $t_{\rm{end}}$ in the following.

The field-space directions orthogonal to $\hat{\sigma}^I$ are
given by
\begin{align}
\hat{s}^{IJ} = G^{IJ}-\hat{\sigma}^I \hat{\sigma}^J,
\end{align}
and $\hat{\sigma}^I$ and $\hat{s}^{IJ}$ tensors are related by
relations~\cite{DeCross:2015uza}
\begin{align}
&\hat{\sigma}^I \hat{\sigma}_I = 1,~~~\hat{s}^{IJ}\hat{s}_{IJ} = N-1, ~~~\hat{\sigma}_I \hat{s}^{IJ} = 0~\text{for~each~}J\label{ortho}.
\end{align}
$N=2$, and $I,J=1,2$ in our two-field scenario.  We can now decompose
the perturbations in the directions of $\hat{\sigma}^I$ and
$\hat{s}^{IJ}$ as
\begin{align}
&Q_\sigma = \hat{\sigma}_I Q^I \label{eq:Qsig},\\
&\delta s^I  =  \tensor{\hat{s}}{^I_J} Q^J\label{eq:deltas},
\end{align}
with $Q_\sigma$ and $\delta s^I$ being referred to as adiabatic and
entropy perturbations, respectively.
We also define a ``turning vector'' $\omega^I$ as the covariant rate of change
of $\hat{\sigma}^I$,
\begin{align}
\omega^I =  \mathcal{D}_t  \hat{\sigma}^I.
\end{align}
The turning vector is orthogonal with respect to $\hat{\sigma}^I$,
$\omega_I\hat{\sigma}^I=0$, the corresponding unit vector is
\begin{align}
\hat{\omega}^I =\frac{\omega^I}{\omega},
\end{align}
with $\omega= |\omega^I|=\sqrt{G_{IJ} \omega^I\omega^J}$. 

With these definitions in place, we can now define the entropy perturbations as
\begin{align}
&Q_s = \hat{\omega}_I Q^I,
\end{align}
which are conveniently normalized to give
\begin{align}
\mathcal{S} = \frac{ H}{\dot{\sigma}} Q_s.\label{eq:entropy1}
\end{align}
The gauge-invariant curvature (adiabatic) perturbation
$\mathcal{R}$~\cite{Mukhanov:1990me,Bassett:2005xm,Malik:2008im}
\begin{align}
\mathcal{R} = \psi -\frac{ H}{\rho+p} \delta q,\label{gaugeinvR}
\end{align}
with $\rho$, $p$ as defined above, and $\delta q$ given by $\partial_i \delta q = -T_{0i}$ evaluated at background order (cf. Appendix~\ref{energymomtensor}) together with Eqs.~\eqref{deltaphiI} and~\eqref{mukh-sasaki}
\begin{align}
\delta q = - G_{IJ} \dot{\varphi}^I \delta\phi^J = -\dot{\sigma}\hat{\sigma}_I \left(Q^I - {\dot \varphi^I \over H}\psi \right).
\end{align}
Therefore, $\mathcal{R}$ takes the compact form
\begin{align}
\mathcal{R} = 
\psi + \frac{ H}{\dot\sigma^2} \dot{\sigma}\hat{\sigma}_I \left(Q^I - \dot{\sigma}{\hat \sigma^I\over H}\psi \right)=
\frac{ H}{\dot{\sigma}} Q_\sigma\label{eq:curvpurt},
\end{align}
at linear order. In the presence of entropy perturbations, the gauge-invariant curvature perturbation does not need to be conserved, $\dot{\mathcal{R}}\neq 0$. The non-adiabatic pressure perturbation is given by \cite{Bassett:2005xm,Malik:2008im}
\begin{align}
\delta p_{\text{nad}} = \delta p - \frac{\dot{p}}{\dot{\rho}}\delta\rho = -\frac{2 \hat\sigma^I \partial_IV }{3 H\dot{\sigma}}\epsilon_m + 
2\dot{\sigma} \left(\omega_I \delta s^I\right).
\end{align}
with $\epsilon_m$ as the comoving density perturbation. For super-horizon scales $k \ll a H$, the only source of non-adiabatic pressure stems from $\delta s^I$.  This means that $\dot{\mathcal{R}}\neq 0$ will not vanish even at the $k \ll a H$ scale and $\omega_I \delta s^I$ will source $Q_\sigma$ and hence $\dot{\mathcal{R}}$.

The gauge invariant curvature perturbation is defined
as~\cite{Mukhanov:1990me,Bassett:2005xm}
\begin{align}
\langle\mathcal{R}(\bm{k}_1) \mathcal{R}(\bm{k}_2) \rangle= (2\pi)^3 \delta^{(3)}(\bm{k}_1+\bm{k}_2) P_{\mathcal{R}}(t;k_1)
\end{align}
and $P_{\mathcal{R}}(t;k)= |\mathcal{R}|^2$. The dimensionless power
spectrum for the adiabatic perturbation is given by
\begin{align}
\mathcal{P}_{\mathcal{R}}(t;k)= \frac{k^3}{2\pi^2}|\mathcal{R}|^2\label{eq:powadia}. 
\end{align}
Similarly, the power spectrum for the entropy perturbations is
\begin{align}
&\mathcal{P}_{\mathcal{S}}(t;k)= \frac{k^3}{2\pi^2}|\mathcal{S}|^2\label{eq:powentrop}.
\end{align}
To find the power spectra of the curvature and isocurvature (entropy) perturbations, Eqs.~\eqref{eq:powadia} 
and~\eqref{eq:powentrop}, we utilize the quantities $H$, $\epsilon$ and unit vectors such as
$\hat{\sigma}^I$, $\hat{\omega}^I$,\dots, from the solutions of the
Eqs.~\eqref{hubble1} and \eqref{eq:bkg} while $Q_\sigma$ and $Q_s$ are
evaluated using the solutions of mode equations from
Eq.~\eqref{eq:fluc}. For a given Fourier mode $k$, we calculate the
different power spectra at the $t=t_{\text{end}}$ numerically as a
function of $k$ as
\begin{equation}
\mathcal{P}_{\mathcal{R}}(k)=\mathcal{P}_{\mathcal{R}}(t_{\text{end}};k),\hspace{2cm}
\mathcal{P}_{\mathcal{S}}(k)=\mathcal{P}_{\mathcal{S}}(t_{\text{end}};k),
 \end{equation}
where $t_{\text{end}}$ denotes the time when inflation ends, i.e.~when $\epsilon =1$.

The spectral index $n_{s}$ of the power spectrum of the curvature
perturbation is defined as
\begin{align}
n_{s} = 1 + \frac{d\ln\mathcal{P}_{\mathcal{R}}(k)}{d\ln k}\label{specin1}. 
\end{align}
As we will discuss in the next section, although our scenario involves
scalar fields $h$ and $\phi$, we shall primarily focus on a scenario
where the dynamics are essentially described by single field-like
inflation. In such a case, the spectral index can be calculated as
\begin{align}
n_{s}(t_\ast) \approx 1 -6 \epsilon(t_\ast) + 2 \eta(t_\ast) \label{specin2},
\end{align}
where $t_\ast$ denotes the time when the reference scale exited the
horizon and the tensor-to-scalar ratio is given by
$r\approx16 \epsilon$.

\begin{table}[!t]
\begin{tabular}{|c |c| c| c| c | c | c| c | c |c| c| c| c}
    \hline
	BP                  & $\xi_R$             &  $\xi_H$   &  $\varphi(t_{\text{in}})$ [$M_{\rm P}$]  & $h_0(t_{\text{in}})$  [$M_{\rm P}$] \\
   \hline
        $a$                   & $2.35\times 10^9$   &  $10^{-3}$ &  5.5                                  & $2\times10^{-4}$ \\
        $b$                   & $2.55\times 10^9$   &  $1$       &  5.5                                  & $8.94\times10^{-4}$ \\
        $c$                   & $2.2\times 10^9$   &   $10$        &  5.4                                  & $5.00\times10^{-3}$ \\
	\hline
	\end{tabular}
	\caption{Benchmark points chosen for our analysis. Scales are given in units of the Planck mass~$M_{\rm P}$. See text for details.}
	\label{parmeterchoices}
\end{table}

We choose three benchmark points to highlight quantitatively the 
implications of consistent inflation parameter choices when 
contextualized with baryogenesis. These are summarized in
Tab.~\ref{parmeterchoices} alongside the required initial field
values to satisfy Planck 2018 measurements: At the pivot scale
$ k = k_{*} $, the amplitude of $\mathcal{P}_{\mathcal{R}}(k)$ should
match the scalar amplitude measurement of~Ref.~\cite{Planck:2018jri}
\begin{equation}
A_{s} = (2.099\pm 0.014)\times 10^{-9}~\text{at 68\% CL}.
\end{equation} 
As a guideline for our parameter choices and the initial values of the
background fields, we follow the valley approximation that we discuss in
Appendix~\ref{valleyapprox}. We note that, whilst finding the parameter
sets, we also ensure that the isocurvature mode remains orders of
magnitude smaller than the curvature perturbation. The background
equations are solved with initial conditions $\varphi(t_{\text{in}})$
and $h_0(t_{\text{in}})$ as in Tab.~\ref{parmeterchoices}, with
vanishing time derivatives; $t_{\text{in}}$ denotes the initial time
for our numerical analysis in the following. The perturbation
equations~\eqref{eq:fluc} are solved with approximate initial
conditions for a Fourier mode $k$
\begin{align}
&Q^I(t)  \simeq \frac{H}{\sqrt{2 k^3}}\bigg(i+ \frac{k}{a H}\bigg)\exp\left\{{i\frac{k}{a H}}\right\}\label{inifluc1},
\end{align}
sufficiently in the past such that the Hubble parameter at $t_\text{in}$ remains approximately constant. In practice, we initialize the $Q^I$ and their
derivatives about four $e$-foldings before they exit the horizon for each mode.

\begin{figure}[!t]
\centering
\includegraphics[height = 4.9cm]{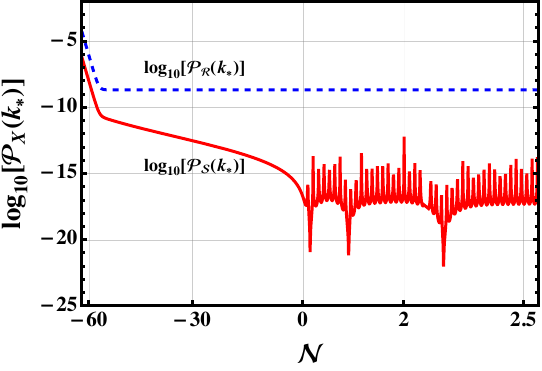}\hspace{5mm}
\includegraphics[height = 5.05cm]{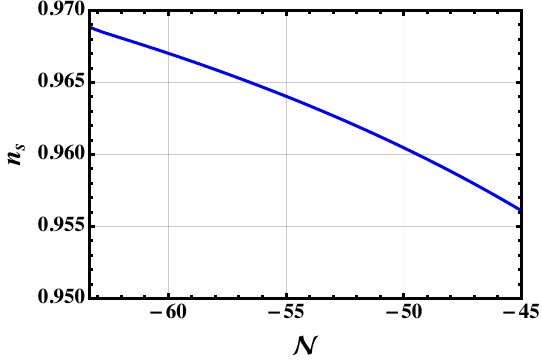}
\caption{The power spectra of the adiabatic and isocurvature modes and the spectral index $n_s$ for the parameter values of BP$a$ in Tab.~\ref{parmeterchoices}.}
\label{powspec}
\end{figure}

\begin{figure}[!t]
\centering
\includegraphics[height = 4.6cm]{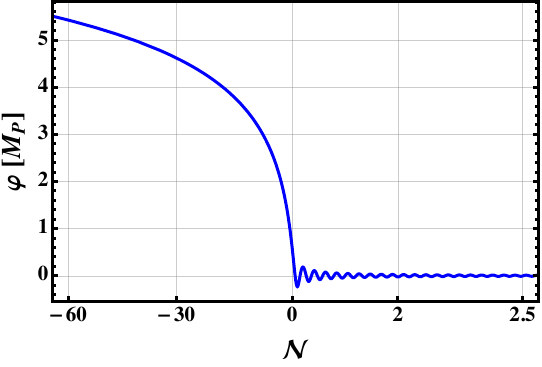}\hspace{1.5cm}
\includegraphics[height = 4.6cm]{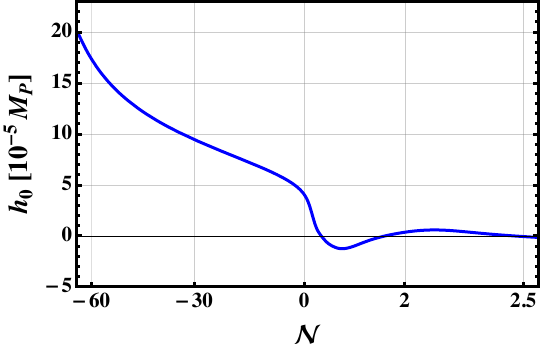}
\caption{The evolution of the background fields $\varphi$ and the $h_0$ for the parameter values of BP$a$ in Tab.~\ref{parmeterchoices}.}
\label{field-evolution}
\end{figure}

\begin{figure}[!t]
\centering
\includegraphics[height = 4.6cm]{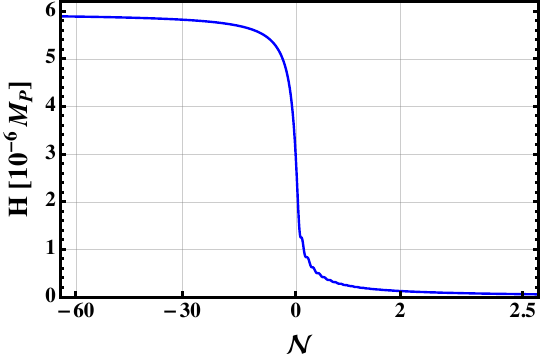}\hspace{1.5cm}
\includegraphics[height = 4.6cm]{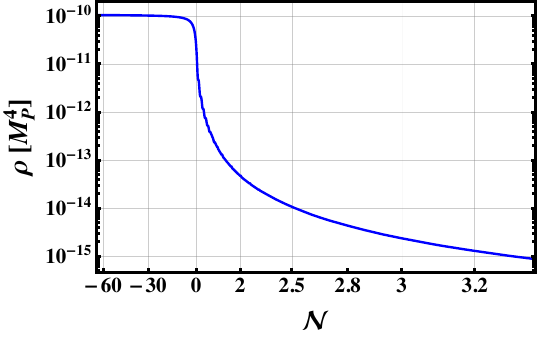}
\caption{The  Hubble function $H$ and the inflaton energy density $\rho$ as in Eq.~\eqref{eq:infenergy} for the parameter values of BP$a$ given 
in Tab.~\ref{parmeterchoices}.}
\label{hubbleenergy0}
\end{figure}

In Fig.~\ref{powspec}, we show the evolution of power spectra
$\mathcal{P}_{\mathcal{R}}$ and $\mathcal{P}_{\mathcal{S}}$ (for the
pivot scale $k = k_*$) and the spectral index $n_s$ for BP$a$. Note,
when calculating both power spectra, we solve Eq.~\eqref{eq:powadia}
and Eq.~\eqref{eq:powentrop} numerically without any assumption
related to slow-roll. It is clear from Fig.~\ref{powspec} that the
isocurvature mode is orders of magnitude smaller than the adiabatic
mode and both power spectra freeze out once they exit the horizon. We
remark that while finding the power spectrum we always check the
orthogonality conditions of Eq.~\eqref{ortho} in our numerical
analysis. In the following, we interchangeably use the cosmological
time $t$ and the number of $e$-foldings before the end of inflation
which is defined as
\begin{align}
\mathcal{N} \equiv \ln \frac{a(t)}{a(t_{\rm{end}})}.
\end{align}
The pivot scale $k_*$ exits the horizon $\mathcal{N_*}=$ 57, 59.3,
54.9 $e$-foldings before the end of inflation for BP$a$, BP$b$, and
BP$c$, respectively.  For illustration, we also show the fields' time
evolution in Fig.~\ref{field-evolution} for BP$a$, while, the
evolution of the Hubble parameter and the inflaton energy density are
shown in Fig.~\ref{hubbleenergy0}.  It is also clear from
Fig.~\ref{powspec} that the spectral index $n_{s}$ lies within the
Planck 2018 range when the reference scale exits the horizon. The
corresponding values of the tensor-to-scalar ratios are
$r_*\sim 0.003$, which is consistent with expected values for
$R^2$-Higgs inflation.

\section{Gauge Field Production}\label{sec:magneto}
The EoM for the gauge field $A_\mu$ of Eq.~\eqref{eq:Bmu-broken} can be rewritten as
\begin{equation} \begin{aligned}
 & \frac{1}{\sqrt{-g_E}}\partial_\alpha \left(\sqrt{-g_E} g_E^{\mu \alpha} g_E^{\nu \beta} F_{A\mu\nu} \right)
+\frac{8 \cos^2\theta_W M_{\rm P}^2}{\xi_R \Lambda^2} \partial_\alpha\left( F(\phi^I)e^{\sqrt{\frac{2}{3}}\, \frac{\phi}{M_{\rm P}}} \right)\tilde{F}_{A}^{\alpha\beta} \\
&\hspace{7cm} - ie Q_f\, e^{-\sqrt{\frac{2}{3}}\frac{\phi}{M_{\rm P}}} g_E^{\mu\beta} \bar f e^a_\mu\tilde{\gamma}_a f=0,
\end{aligned} \end{equation}
without the presence of a torsion term $F_{A\mu\nu} = D_\mu A_\nu - D_\nu A_\mu = \partial_\mu A_\nu - \partial_\nu A_\mu$.
One can identify  the fermion current
\begin{equation} 
j^\mu=\sum_f  ie Q_f\,  e^{-\sqrt{\frac{2}{3}}\frac{\phi}{M_{\rm P}}} g_E^{\mu\nu} \bar f e^a_\nu\tilde{\gamma}_a f \label{eq:fermcurr}
\end{equation}
that sources the Schwinger effect.

\subsection*{Neglecting the Schwinger effect}\label{sec:magnetowoSch}
First, we consider the scenario without Schwinger effect i.e.~when the fermion current is negligible. This is possible if the fermion field values 
are small. One can now separate the space and time component of the $A_\mu$ field. The time component of Eq.~\eqref{eq:Bmu} at linear order in the perturbations is
\begin{align}
-\frac{1}{a^2}\partial_i\left(\partial_i A_0 -\partial_0 A_i\right) = 0,\label{eq:0A1}
\end{align}
which, in temporal gauge $A_0 = 0$, reduces to $\partial_i\dot{A}_i = 0$.
The spatial components of Eq.~\eqref{eq:Bmu} are found to be
\begin{equation}
\ddot{A}_i +H\dot{A}_i
-\frac{1}{a^2}\partial_j \left(\partial_j A_i - \partial_i A_j\right) 
-\frac{\xi H}{a}\epsilon^{i jk}
\left(\partial_j A_k - \partial_k A_j\right)=0.
\label{eq:iA}
\end{equation}
where
\begin{equation}
H= \frac{\dot{a}}{a},  \hspace{2cm} \xi = \frac{4\cos^2\theta_W M_{\rm P}^2}{\xi_R \Lambda^2 H} \;\partial_0\left( F(\varphi^I)e^{\sqrt{\frac{2}{3}}\, \frac{\varphi}{M_{\rm P}}} \right). \label{def:xi}
\end{equation}
In momentum space, using the notation $\vb A\equiv \vec A$
\begin{equation}
\vb A(t,\vb x) = \int \frac{d^3\vb{k}}{\left(2\pi\right)^{3/2}} \tilde{\vb A}(t,\vb{k}) e^{-i \vb{k}\cdot \vb{x}},\label{eq:Aifou}
\end{equation}
with $|\vb{k}|=k$, Eq.~\eqref{eq:iA} reads
\begin{align}
& \ddot{\tilde{\vb{A}}} + H \dot{\tilde{\vb{A}}}+ \frac{k^2}{a^2} \tilde{\vb{A}} +\frac{2i\xi H}{a} 
(\vb{k \times \tilde{A}})  = 0 .
\label{photon:equation}
\end{align}
The $\tilde{\vb{A}}$ field can be written in terms of transverse components as 
\begin{align}
\tilde{\vb{A}} =  \sum_{\lambda=\pm} \tilde{A}^\lambda(t,\vb{k})\ \hat{\epsilon}^\lambda(\vb{k}),~\text{with}~\vb{k}\cdot\hat{\epsilon}^\lambda(\vb{k}) = 0,~~ i\vb{k}\times\hat{\epsilon}^\lambda(\vb{k}) = \lambda k \ \hat{\epsilon}^\lambda(\vb{k}) \label{photon:pol}.
\end{align}
so that, using conformal time $\tau$ (with $\partial_0 = \partial_t = {a}^{-1} \partial_\tau$), the EoM for the transverse components becomes
\begin{align}
\partial_\tau^2  \tilde{A}^\lambda +\omega_\lambda^2   \tilde{A}^\lambda  = 0, \label{eq:transphoton}
\end{align}
with
\begin{align}
\omega_\lambda^2(\tau,k)= k^2  +
2 \lambda\xi Ha k. \label{eq:photontransmode}
\end{align}
In order to quantize the gauge fields, we first integrate
Eq.~\eqref{eq:transphoton} by parts to get the action quadratic in the
fields
\begin{align}
S^\lambda = \int d\tau \,\mathcal{L}_\lambda = \int \, d\tau \,d^3\vb{k}\, \bigg[\frac{1}{2} |\partial_\tau  \tilde{A}^\lambda|^2
- \frac{1}{2}\omega_\lambda^2(\tau,k) |\tilde{A}^\lambda|^2\bigg].
\end{align}
As we deal with non-canonical kinetic terms, we
apply the quantization procedure detailed in Ref.~\cite{Lozanov:2016pac}. The canonical momentum of the transverse modes are 
\begin{align}
\pi_\lambda(\tau,\vb{k})=\frac{\delta \mathcal{L}_\lambda}{\delta\left(\partial_\tau  \tilde{A}^\lambda(\tau,-\vb{k})\right)},
\end{align}
with the commutation relation expressed as
\begin{align}
\bigg[ \tilde{A}^\lambda(\tau,\vb{k}), \partial_\tau \tilde{A}^{\lambda'}(\tau,\vb{q})\bigg]= i \delta_{\lambda\lambda'}\delta(\vb{k}+\vb{q}).
\end{align}
The field operator $\tilde{A}^\lambda(\tau,\vb{k})$ can be written as
creation and annihilation operators
\begin{align}
\tilde{A}^\lambda(\tau,\vb{k}) = \hat{a}_{\vb{k}}^\lambda u_k^\lambda(\tau) + \hat{a}_{\vb{-k}}^\lambda u_k^{\lambda\*}(\tau),
\end{align}
and the mode equations for the gauge fields are then
\begin{align}
&\ddot{u}^\lambda + H \dot{u}^\lambda+
\frac{\omega_\lambda^2}{a^2} \,u^\lambda =0.
\label{eq:modephoton}
\end{align}
From these mode functions, we can compute the gauge observables, namely the magnetic and electric fields' energy densities, magnetic helicity and its derivative, defined as
\begin{subequations} \begin{eqnarray}
\rho_B &=& \frac{1}{a^4} \int_{k_{\text{min}}}^{k_c} dk \frac{k^4}{4\pi^2}\sum_\lambda |u^\lambda|^2,\label{eq:energydenB}\\
\rho_E &=& \frac{1}{a^4} \int_{k_{\text{min}}}^{k_c} dk \frac{k^2}{4\pi^2}\sum_\lambda |\partial_\tau u^\lambda|^2,\label{eq:energydenE}\\
\mathcal{H} &=&  \frac{1}{a^3} \int_{k_{\text{min}}}^{k_c} dk \frac{k^3}{2\pi^2} \left(|u^+|^2-|u^-|^2\right),\label{eq:hyperheli}\\
\mathcal{G} &=& \frac{1}{2} \frac{\partial \mathcal{H}}{\partial t},\label{eq:hyperhelideri}
\end{eqnarray} \label{eq:plasma-observables} \end{subequations}
with the cut-off value given by~\cite{Gorbar:2021rlt,Cado:2022pxk}
\begin{align}
k_c = 2 \left| aH\,\xi \right|,\label{eq:cutoffk}
\end{align}
defined by the condition $\omega_\lambda^2(\tau,k_c)=0$ satisfied by the helicity $\lambda$ such that $\textrm{sign}(\lambda\xi)=-1$.
The corresponding $U(1)_Y$ quantities are linked to the electromagnetic ones via
\begin{equation} 
\begin{aligned}
\rho_{B_Y} = \rho_{B} \;\cos^2{\theta_W} \hspace{3cm} &\rho_{E_Y} =  \rho_{E}\;\cos^2{\theta_W}  \\  \mathcal{H}_{Y} = \mathcal{H}\;\cos^2{\theta_W}  \hspace{3.3cm}& \mathcal{G}_{Y} = \mathcal{G}\;\cos^2{\theta_W} 
\end{aligned} 
\end{equation}
In general, the integration limits should cover all modes from zero to infinity, however, not all modes are amplified during inflation. At the time $t$, the cut-off mode $k_c$ is found by the solution of $\omega_\lambda=0$; 
essentially this is when a mode $k = k_c$ crosses the horizon for the first time (at least for one helicity). The modes $k \gg k_c$ are not excited during inflation and can be
neglected for the estimation of the above observable quantities. We will discuss $k_{\rm{min}}$ shortly.

In order to find $\rho_B$, $\rho_E$, $\mathcal{H}$ and $\mathcal{G}$ we solve Eq.~\eqref{eq:modephoton} numerically via fourth-order Runge-Kutta (RK4) method in discrete time steps. We outline the details in Appendix~\ref{rk4method}. For the $i$-th time step, the gauge field modes are initialized with the Bunch-Davies (BD) initial condition as~\cite{Cuissa:2018oiw}
\begin{align}
u_\lambda(k,t_i) \simeq \frac{1}{\sqrt{2 k}} e^{-i \omega_i t_i}, \hspace{2.5cm}
\dot{u}_\lambda \simeq -i \frac{\omega_i}{\sqrt{2 k}} e^{-i \omega_i t_i},\label{eq:ini}
\end{align}
with $\omega_i = {k}{a^{-1}(t_i)}$. It is practically not possible to go to the infinite past. Hence, to ensure that all modes remain well within the horizon at the initial time step~$t_{\rm{in}}$, we chose $k_{\rm{min}} = x_{\rm{BD}}a(t_{\rm{in}}) H(t_{\rm{in}})$ with $x_{\rm{BD}}=100$. On the one hand, if a mode $k$ remains well within the horizon, $k > x_{\rm{BD}}\,a(t_i) H(t_i)$, we directly assume the BD solutions for the modes instead of applying the RK4 method for any subsequent time step. On the other hand, all superhorizon modes are solved with the RK4 method. For the numerical solution discussed below, we employ 25k time steps.

\begin{figure}[!t]
\centering
\includegraphics[width=.44 \textwidth]{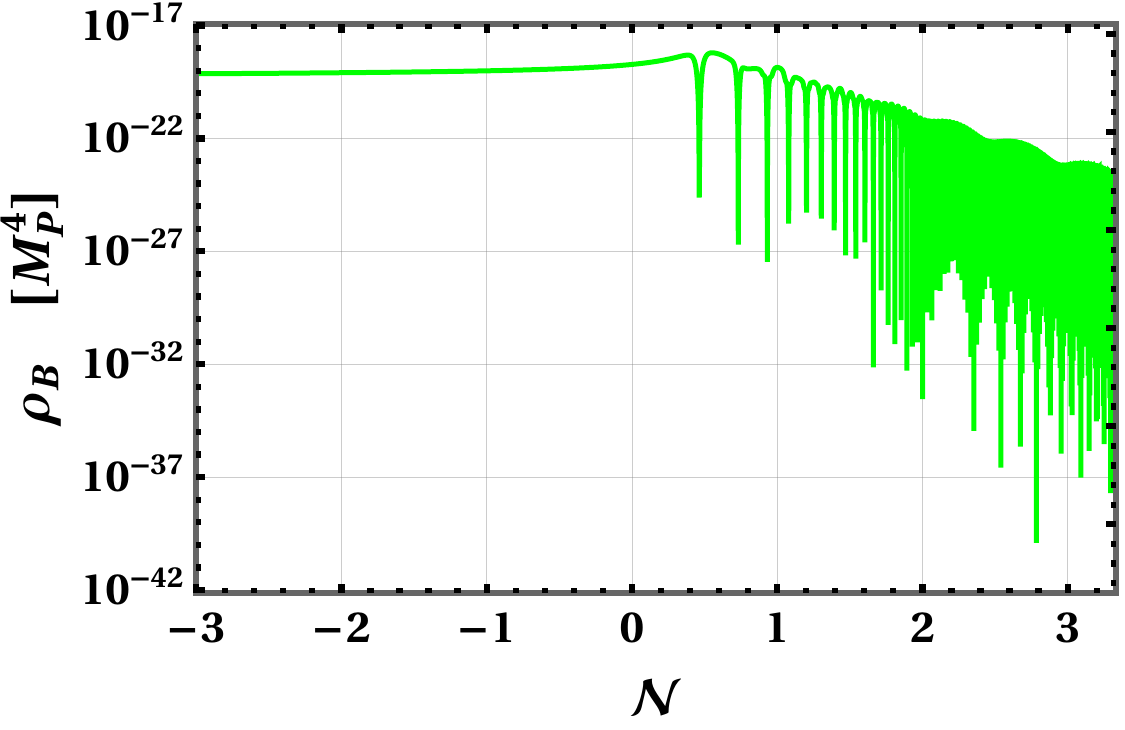}\hspace{0.9cm}
\includegraphics[width=.44 \textwidth]{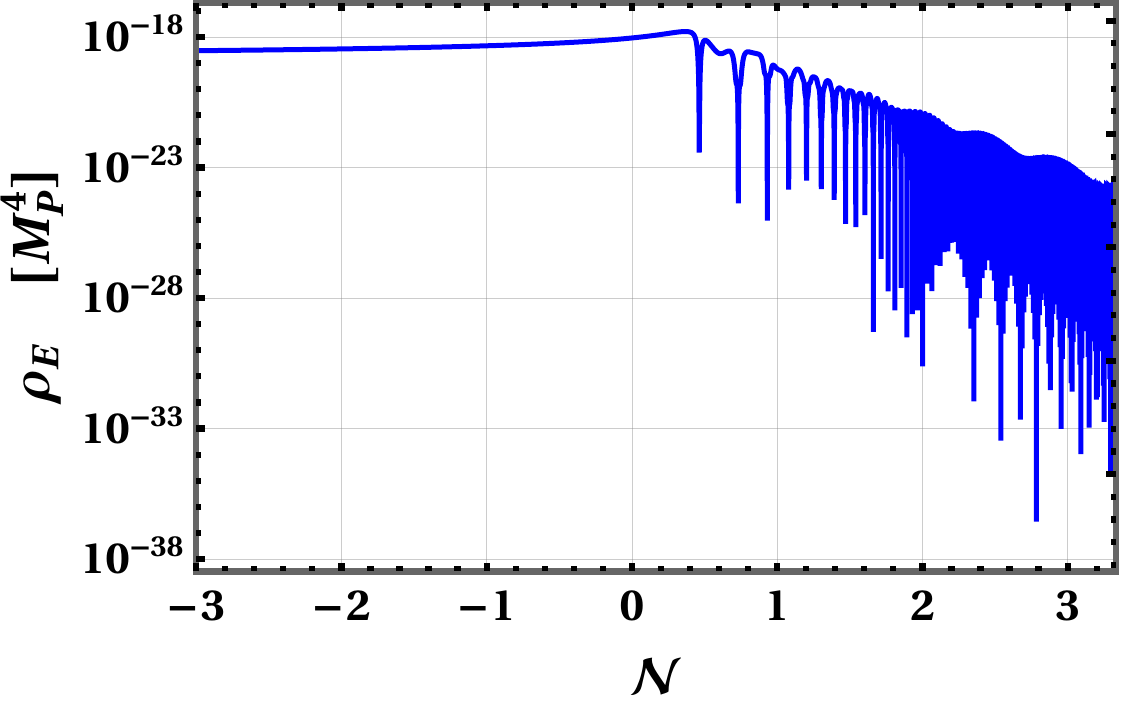}\\[0.5cm]
\includegraphics[width=.44 \textwidth]{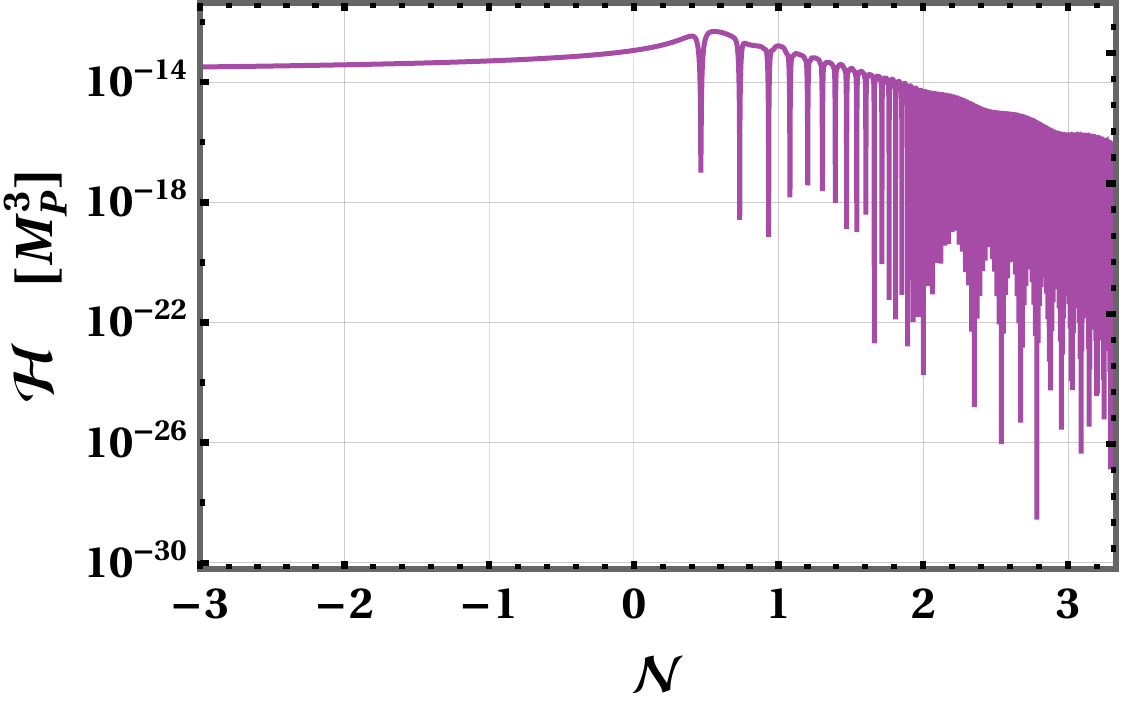}\hspace{0.9cm}
\includegraphics[width=.44 \textwidth]{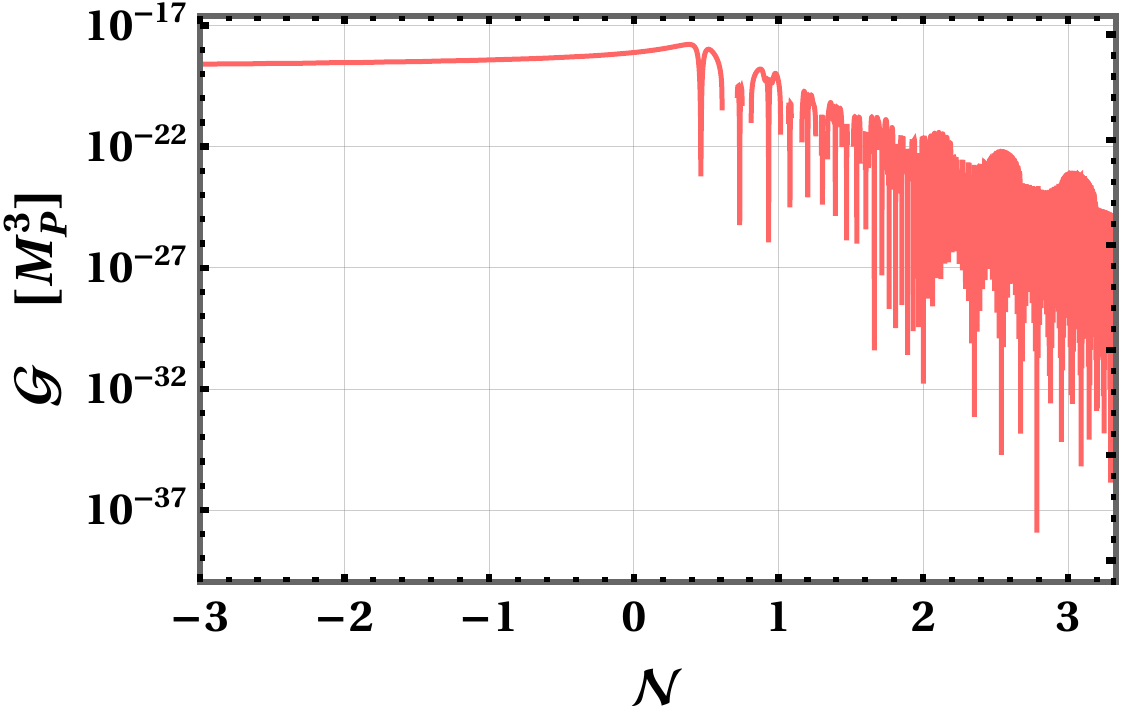}
\caption{In the upper panel we plot the energy densities $\rho_B$ and $\rho_E$ with $\Lambda=2.55\times 10^{-5}$~$M_{\rm P}$ for the BP$a$ summarized in  Tab.~\ref{parmeterchoices}. The lower panel corresponds to the hyperhelical magnetic fields $\mathcal{H}$ and  $\mathcal{G}$ for identical parameter choices.}
\label{energydens}
\end{figure}

In Fig.~\ref{energydens}, we show the evolution of $\rho_B$, $\rho_E$, $\mathcal{H}$ and $\mathcal{G}$ for BP$a$ with $\Lambda= 2.55\times10^{-5}M_{\rm P}$ for illustration. Similar values are found for the other benchmark points.  We remark that we have compared our numerical results to the analytical approximation of the magnetic and electric fields' energy densities, magnetic helicity, and its derivative as in Ref.~\cite{Cado:2022pxk} and find good agreement.

\subsection*{Relevance of the Schwinger effect}\label{sec:magnetoSch}
We now turn to the impact of the Schwinger effect. The fermion current of Eq.~\eqref{eq:fermcurr} can be expressed as
\begin{equation} 
j^\mu=(\rho_{\rm c},\bm{J})
\end{equation}
The current and the gauge field are related by Ohm's law
\begin{equation}
\bm{J} = \sigma_{\rm c} \,\bm{E} = -  \sigma_{\rm c}\,\partial_\tau \bm{A},\label{Ohm-law}
\end{equation}
where the conductivity $\sigma_{\rm c}$ has been defined as a comoving quantity. The physical conductivity $\sigma_{\rm ph}$ relates to the comoving one via $\sigma_{\rm c}=a\,\sigma_{\rm ph}$.
In the case of one Dirac fermion $f$ with mass $m_f$ and charge $Q_f$ under a $U(1)$ group with coupling $g$, the comoving conductivity associated to $f$ can be written as~\cite{Domcke:2018eki}
\begin{equation}
\sigma_f^{\rm c}= \frac{|g \, Q_{f} |^3}{6\pi^2}
\frac{a}{H}\;\sqrt{2\rho_B}\;\coth\left(\pi\sqrt{\frac{\rho_B}{\rho_E}} \right)\exp{-\frac{\pi m_f^2}{\sqrt{2\rho_E}\, |g\,Q_{f}|}-\sqrt{\frac{2}{3}}\frac{\varphi}{M_{\rm P}}},
\label{eq:sigma-def}
\end{equation}
where $m_f\equiv m_f(h_0)= m_f(v)h_0/v$ and
so that
\begin{equation}
\sigma_{\rm c}=\sum_f \sigma^{\rm c}_f=\sum_\ell \sigma^{\rm c}_\ell+N_c\sum_q\sigma^c_q,
\end{equation}
with $\ell=e,\mu,\tau$; $q=u,d,c,s,t,b$, $N_c=3$ being the number of colors. Last, since we are in the broken phase, we identify $g$ as the electric charge $g\equiv e\simeq 0.71$ at the scale in which inflation takes place.

This conductivity is to be distinguished from the conductivity of a thermal plasma after reheating in a radiation-dominated universe. We stress that the above is the conductivity at the end of inflation, before the reheating, produced by fermion pair formation from the magnetic field.  Also, this estimation is valid in the case of collinear electric and magnetic fields, an assumption that we have numerically checked. Finally, the electric and magnetic fields are assumed to be slowly varying, as we expect the hypercharge gauge field to reach a stationary configuration, where the tachyonic instability and the induced current balance each other. We have verified in our numerical simulation that this is indeed the case. 

In the presence of the fermion current, Eq.~\eqref{photon:equation} becomes
\begin{align}
& \ddot{\tilde{\vb{A}}} + (H+\sigma_{\rm ph})\, \dot{\tilde{\vb{A}}}+ \frac{k^2}{a^2} \;\tilde{\vb{A}} +\frac{2iH\xi}{a}\;
(\vb{k \times \tilde{A}})  = 0,
\label{photon:equation_sch}
\end{align}
which, for the transverse components in conformal time reads as
\begin{equation}
\partial_\tau^2  \tilde{A}^\lambda +\sigma_{\rm c} \, \partial_\tau \tilde{A}^\lambda+\omega_\lambda^2   \tilde{A}^\lambda  = 0,
\end{equation}
which can be recast as 
\begin{align}
\partial_\tau^2  \tilde{A}^\lambda + \left(\frac{\partial}{\partial \tau}\log(\Delta(\tau)) \right) \partial_\tau \tilde{A}^\lambda+\omega_\lambda^2   \tilde{A}^\lambda  = 0, \label{eq:transphoton_sch}
\end{align}
with
\begin{align}
\Delta(\tau) = \exp{\int_{-\infty}^{\tau}\sigma_{\rm c}(\tau')\,d\tau'} .
\end{align}
Integrating Eq.~\eqref{eq:transphoton_sch} by parts as in the previous subsection, one can now define the canonical momentum for the transverse modes as 
%
\begin{figure}[h]
\centering
\includegraphics[height = 5cm]{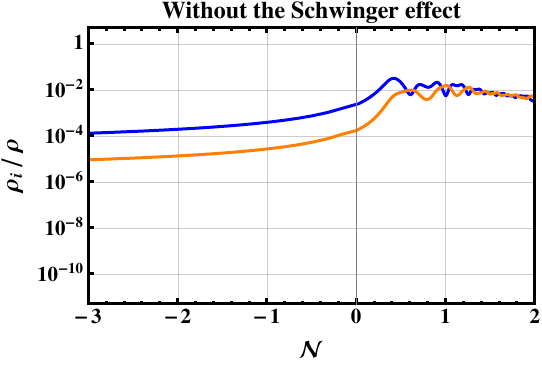}\hspace{4mm}
\includegraphics[height = 5cm]{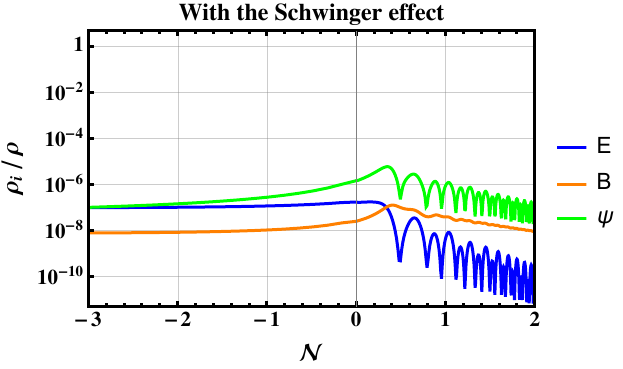}
\caption{Energy breakdown for $\xi_R=2.35\times 10^9$, $\xi_H=10^{-3}$ and $\Lambda= 2\times 10^{-5}M_{\rm P}$ at the end of inflation and the onset of reheating. We show a comparison between the absence and the presence of the Schwinger effect for the quantities $\rho_i (\mathcal{N}) / \rho (\mathcal{N})$, $i=E,B,\psi$. When the Schwinger effect is strong, like here, the fermion energy density can dominate over the gauge density. Still, all energy shares are reduced in the presence of the Schwinger effect.
\label{hubbleenergy_breakdown}}
\end{figure}
\begin{align}
&\pi_\lambda(\tau,\vb{k})= \frac{\delta \mathcal{L}_\lambda}{\delta\left(\partial_\tau  \tilde{A}^\lambda(\tau,-\vb{k})\right)} = \Delta(\tau) \partial_\tau \tilde{A}^{\lambda'}(\tau,\vb{k}),
\end{align}
and the commutation relation now becomes~\cite{Lozanov:2016pac} 
\begin{align}
\bigg[ \tilde{A}^\lambda(\tau,\vb{k}), \partial_\tau \tilde{A}^{\lambda'}(\tau,\vb{q})\bigg]= i \frac{1}{\Delta(\tau)}\delta_{\lambda\lambda'}\delta(\vb{k}+\vb{q}),
~~\mbox{with,}~\tilde{A}^\lambda(\tau,\vb{k}) = \hat{a}_{\vb{k}}^\lambda u_k^\lambda(\tau) + \hat{a}_{\vb{-k}}^\lambda u_k^{\lambda\*}(\tau).
\end{align}
The mode equations for the gauge fields in the presence of the Schwinger effect become
\begin{align}
&\ddot{u}^\lambda + (H +\sigma_{\rm c}) \dot{u}^\lambda+
\frac{k}{a}\bigg(\frac{k}{a}  + 
2\lambda H \xi  \bigg)u^\lambda =0,
\label{eq:modephoton_sch}
\end{align}
and the cut-off momenta $k_c$ is now modified to
\begin{align}
k_c =  \left| aH\,\xi \right| + \sqrt{  \left(aH\,\xi \right)^2+ \frac{a^2}{2}  \left[\dot{\sigma}_{\rm ph}+\sigma_{\rm ph}\left(\frac{\sigma_{\rm ph}}{2}+H\right) \right]}\,.\label{eq:cutoffksch}
\end{align}
 At early times solution of the mode equations of Eq.~\eqref{eq:modephoton_sch} are represented by WKB solution~\cite{Lozanov:2016pac,Sfakianakis:2018lzf}
\begin{align}
u^\lambda(\tau) = \frac{1}{\sqrt{2}} \frac{1}{\sqrt{\Delta(\tau) \omega_\lambda(\tau,k)}} \;e^{- i \int_{\tau_{\rm{in}}}^\tau d\tau' \omega_\lambda(\tau',k)}
\end{align}
as long as $|\frac{\partial_\tau \omega_\lambda}{\omega_\lambda}|\ll1$. In practice, we utilize the early-time solution for the modes 
\begin{align}
u^\lambda(\tau) = \frac{1}{\sqrt{2}} \frac{1}{\sqrt{\Delta(\tau) \omega_\lambda(\tau,k)}} \;e^{- i k \tau}
\end{align}
to find relevant, observable quantities. From Eq.~\eqref{eq:modephoton_sch} and Eq.~\eqref{eq:cutoffksch}, we find the energy densities for BP$a$ and display them in the right panel of Fig.~\ref{hubbleenergy_breakdown}. 
 
Due to the coupling between the fermion and gauge sectors, massless hypercharged fermions are continuously produced during inflation. They are massless as long as the EW symmetry remains intact and thus contribute to the energy density of relativistic radiation as
\begin{equation}
 \rho_\psi =  \lim_{V\to\infty}\dfrac{\sigma_{\rm c}}{V}\int_Vd^3x \;\frac{\langle \bm{A} \cdot \bm{E} \rangle}{a^4} =\frac{\sigma_{\rm c}}{a^4}  \int^{k_c}_{k_{\rm min}} d k \, \frac{k^2}{2\pi^2}  \dfrac{d}{d \tau} \sum_\lambda |u^\lambda|^2. \label{def:fermion-energy}
\end{equation}
It has been shown in Ref.~\cite{Gorbar:2021rlt} that the fermion energy density can easily dominate over the energy densities of $E$ and $B$ fields at the end of inflation. This situation has been chosen as an example in Fig.~\ref{hubbleenergy_breakdown} where we display the energy fraction $\rho_i (\mathcal{N}) / \rho (\mathcal{N})$, $i=E,B,\psi$ at the end of inflation and the onset of reheating. We show a direct comparison between the presence and the absence of the Schwinger effect. While for $\Lambda \gtrsim 4 \times 10^{-5}M_{\rm P}$ the difference is an order one factor, the Schwinger effect reduces the amount of electromagnetic energy and helicity up to two orders of magnitude for $\Lambda \simeq 2.4 \times 10^{-5}M_{\rm P}$, see Fig.~\ref{baryogenesis_w_schwinger}. This is because the presence of the Schwinger effect trades an exponential behaviour in $\xi$ with a polynomial one.

When the gauge share dominates at least by 80\%, the Universe will reheat before the perturbative decay of the inflaton~\cite{Cuissa:2018oiw}, a phenomenon called \textit{gauge preheating}. As in Ref.~\cite{Cado:2022pxk}, we found that preheating is unlikely since the ratio is $\sim 10^{-6}$ at most. However, the huge damping in both energy and helicity does not preclude a window in the parameter space where the BAU is achieved, as we will see in the next section.

\section{Baryogenesis}\label{sec:baryogensis}
To generate a baryon asymmetry, the Sakharov conditions~\cite{Sakharov:1967dj} must be met: \textit{(i)} the system must contain a process that violates the baryon number, \textit{(ii)}, this process also violates C/CP symmetries, \textit{(iii)} this process occurs out of thermal equilibrium. In the SM, the CP-violating term from the CKM matrix phase is too small to induce a significant baryon asymmetry at a low energy scale, hence we included the dimension-six CP-odd term between Ricci scalar and $U(1)_Y$ gauge boson. On the other hand, in the symmetric phase of the EW plasma, the SM exhibits a chiral anomaly that is enough to source the present-day BAU. The anomaly expresses the fact that the $B+L$ anomaly, the $U(1)_Y$ helicity and the weak sphaleron are connected as 
\begin{align} 
\Delta N_B = \Delta N_L = N_g \left(  \Delta N_{\rm CS} - \dfrac{g'^2}{16\pi^2}\; \Delta \H_Y\right). 
\label{def:anomaly:vessel}
\end{align}
The factor $N_g=3$ is the number of fermion generations and $g'$ is the $U(1)_Y$ gauge coupling. Under the thermal fluctuation of the $SU(2)_L$ gauge fields, the Chern-Simons number $N_{\rm CS}$ is diffusive, resulting in the rapid washout of both lepton $N_L$ and baryon $N_B$ numbers. On the contrary, a helical primordial magnetic field acts as a source, and a net baryon asymmetry can remain after the EW phase transition. 

In Refs.~\cite{Kamada:2016eeb,Kamada:2016cnb}, the effects of the helicity decay and sphaleron washout balance have been studied within a careful analysis of the transport equations for all SM species during the EWPT. As a result, a non-zero baryon-to-entropy ratio $\eta_B$ remains in the broken phase while the transformation of baryon asymmetry back into helicity is avoided. The novelty of the mechanism lies in the introduction of a time-dependent (temperature-dependent) weak mixing angle $\theta_W(T)$ which enters an additional source of the baryon number into the kinetic equation. When the EW symmetry breaking occurs at $T \simeq 160$~GeV, the primordial hypermagnetic field becomes an electromagnetic field. However, the electroweak sphaleron remains in equilibrium until $T \simeq 130$~GeV and threatens to washout the baryon asymmetry. Therefore proper modeling of the epoch 160~GeV~$\gtrsim T \gtrsim 130$~GeV is critical to an accurate prediction of the relic BAU.

The behavior of $\theta_W(T)$ is confirmed by analytic calculations~\cite{Kajantie:1996qd}, and numerical lattice simulations~\cite{DOnofrio:2015gop}. 
We follow Refs.~\cite{Kamada:2016cnb,Jimenez:2017cdr} and model it with a smooth step function
\begin{align} \cos^2 \theta_W  =  \frac{g^2}{g'^2+g^2}  +  \frac{1}{2}  \frac{g'^2}{g'^2+g^2}\left(1 + \tanh \left[\frac{T - T_{\rm step}}{ \Delta T} \right] \right)  \label{SmoothStepFunction}\end{align}
which, for $155\, \text{GeV}  \lesssim T_{\rm step}  \lesssim 160\, \text{GeV} $ and $5\, \text{GeV} \lesssim \Delta T  \lesssim 20\, \text{GeV} $, describes reasonably well the analytical and lattice results for the temperature dependence.
Consequently, it is possible to generate the observed BAU from a maximally helical magnetic field that was generated before the EW crossover.
Indeed, including all contributions, the Boltzmann equation for the baryon-to-entropy ratio $\eta_B$ reads
\begin{align} \frac{d \eta_B}{d x}\, =\, -\frac{111}{34} \gamma_{W\rm sph} \, \eta_B \, + \, \frac{3}{16 \pi^2} (g'^2 + g^2) \sin(2 \theta_W) \frac{d \theta_W}{d x}  \, \frac{\mathcal{H}_Y}{s} ,\label{BoltzmannEquation}\end{align}
where $x = T/H(T)$, with $H(T)$ being the Hubble rate at temperature $T$, $\mathcal{H}_Y$ the hypermagnetic helicity that is initially present and $s$ the comoving entropy density of the SM plasma given by $s=(2\pi^2/45)g_{\ast} T^3$.
Furthermore, $\gamma_{W \rm sph}=6\,\Gamma_{W \rm sph}/T^4$ is the dimensionless transport coefficient for the EW sphaleron which, for temperatures $T <  161$~GeV, is found from lattice simulations to be~\cite{DOnofrio:2014rug}
\begin{align} \gamma_{W \rm sph} \, \simeq \,  \exp{-147.7 + 107.9 \; \frac{T}{130\, \text{GeV}}}  . \end{align}

The Boltzmann equation \eqref{BoltzmannEquation} has been numerically solved in Ref.~\cite{Kamada:2016cnb} and the baryon-to-entropy ratio $\eta_B$ was found to become frozen, i.e.~$\dot{\eta}_B=0$, at a temperature $T \simeq 135$~GeV. As expected, this is close to the temperature $T \simeq  130$~GeV at which EW sphalerons freeze out. Setting the RHS of Eq.~\eqref{BoltzmannEquation} to zero and solving for $\eta_B$ yields
\begin{align}\eta_B  \simeq   4 \cdot 10^{-12} \, f_{\theta_W}  \frac{\mathcal{H}_Y}{H^3(t_{\text{end}})} \left( \frac{H(t_{\text{end}})}{10^{13} \, \text{GeV}} \right)^{\frac{3}{2}} \left(\frac{T_{\rm rh}}{T_{\rm rh}^{\rm ins}}  \right),\label{constraint-nB} \end{align}
where the (instant) reheating temperature is
\begin{align}T_{\rm rh}= \left(\frac{90}{\pi^2 g_\ast}\right)^{\frac{1}{4}} \sqrt{\Gamma_\phi } ,\hspace{1.5cm}T_{\rm rh}^{\rm ins}= \left(\frac{90}{\pi^2 g_\ast}\right)^{\frac{1}{4}} \sqrt{H(t_{\text{end}})} ,\end{align}
and $\Gamma_\phi$ is the total decay width of the inflaton that reheats the universe after inflation.

All the details on the EWPT dynamics are encoded in the parameter $f_{\theta_W} $ which is subject to significant uncertainties
\begin{align} f_{\theta_W}  = -\sin (2 \theta_W) \, \dfrac{d\theta_W}{d\log T}\bigg\rvert_{T=135\text{ GeV}}, \hspace{1.5cm} 5.6 \cdot 10^{-4} \lesssim f_{\theta_W}  \lesssim 0.32.
\label{eq:ftheta}
\end{align}
The bounds on $f_{\theta_W} $ are given by varying $T_{\rm step}$ and $\Delta T$ in the ranges given below Eq.~\eqref{SmoothStepFunction}.
The result Eq.~\eqref{constraint-nB} is a main ingredient of this work as it directly relates the amount of the final BAU to the amount of hypermagnetic helicity available at the EWPT.

\begin{figure}[htbp]
\centering
\includegraphics[height = 4.8cm]{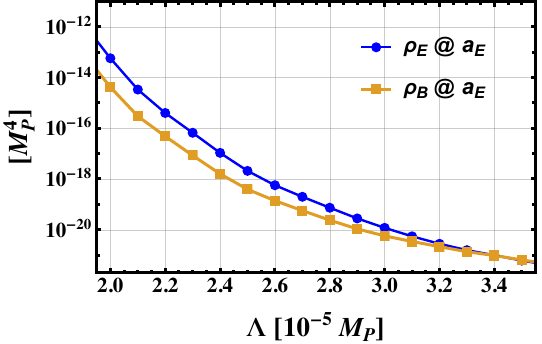}\hspace{13mm}
\includegraphics[height = 4.8cm]{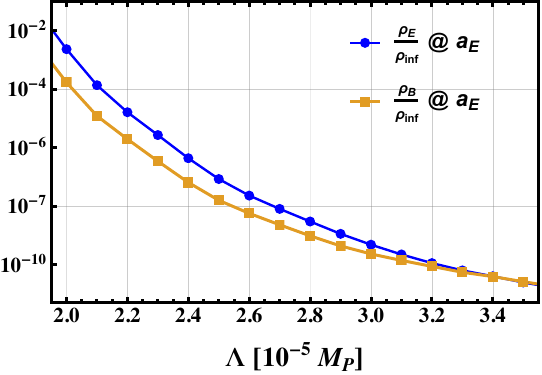}\\\vspace{5mm}
\includegraphics[height = 5cm]{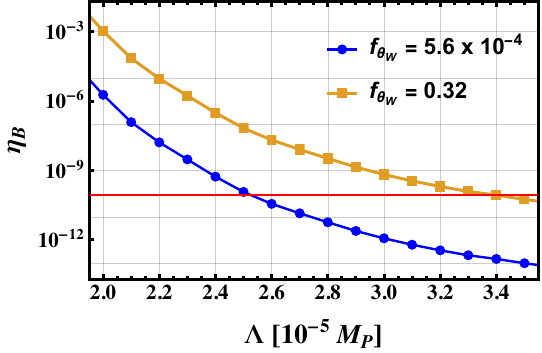}\hspace{13mm}
\includegraphics[height = 5cm]{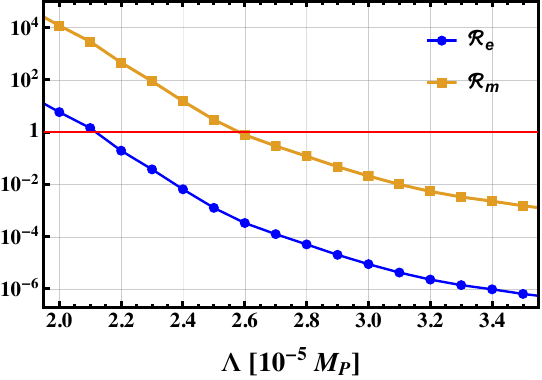}\\\vspace{5mm}
\includegraphics[height = 5cm]{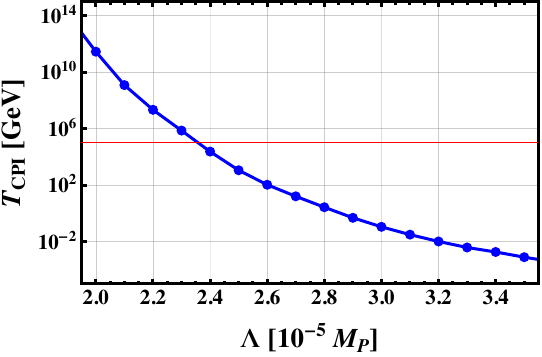}\hspace{9mm}
\includegraphics[height = 5cm]{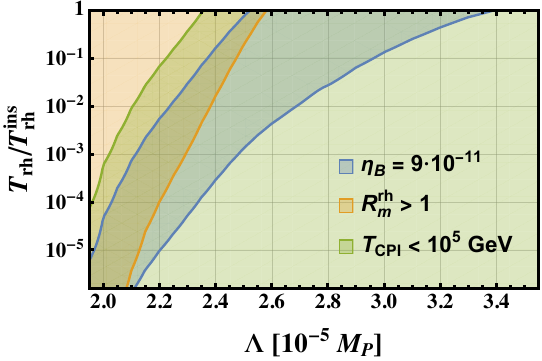}
\caption{These figures display a scan of the parameter $\Lambda$ with the first set of initial conditions, see BP$a$ in Tab.~\ref{parmeterchoices} without Schwinger effect. Top panels: Magnetic and electric energy density (left) and their ratio with the inflation energy density (right). Middle panel: baryon asymmetry $\eta_B$ (left) and Reynolds numbers (right) with their corresponding constraints in red. Bottom panels: CPI temperature with constraint in red (left) and baryogenesis parameter space (right).
}
\label{baryogenesis_wo_schwinger}
\end{figure}

\begin{figure}[htbp]
\centering
\includegraphics[height = 4.8cm]{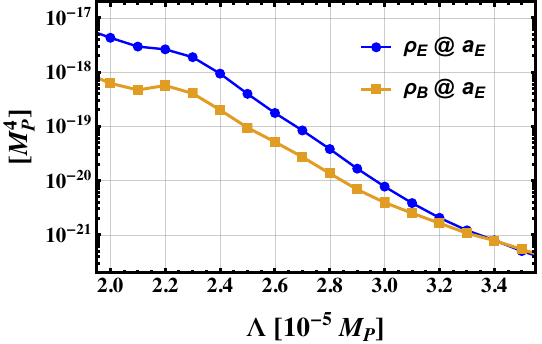}\hspace{13mm}
\includegraphics[height = 4.8cm]{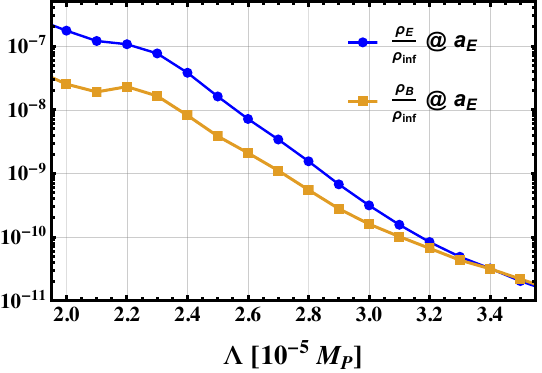}\\\vspace{5mm}
\includegraphics[height = 4.8cm]{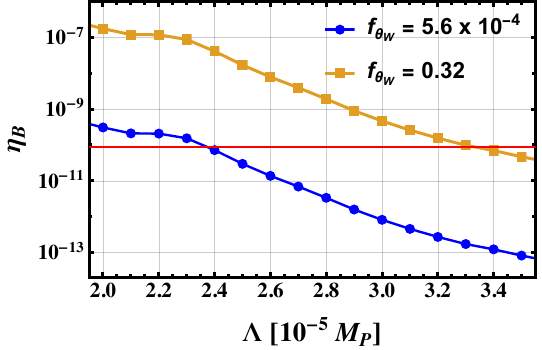}\hspace{13mm}
\includegraphics[height = 4.8cm]{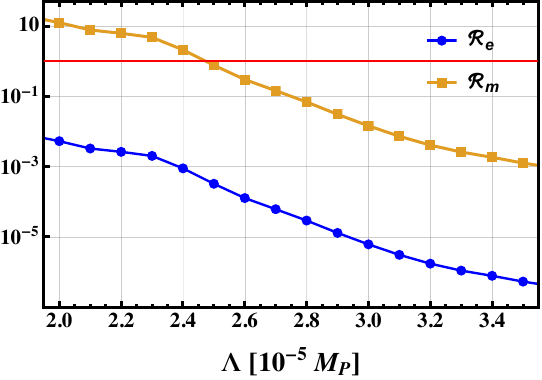}\\\vspace{5mm}
\includegraphics[height = 5cm]{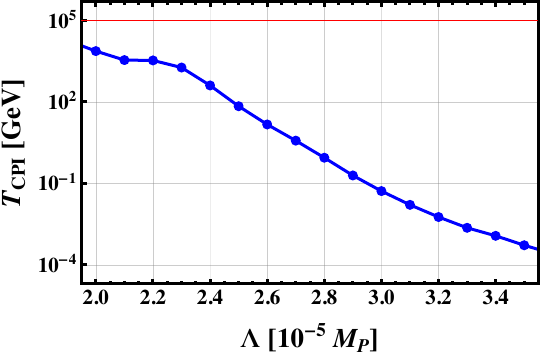}\hspace{9mm}
\includegraphics[height = 5cm]{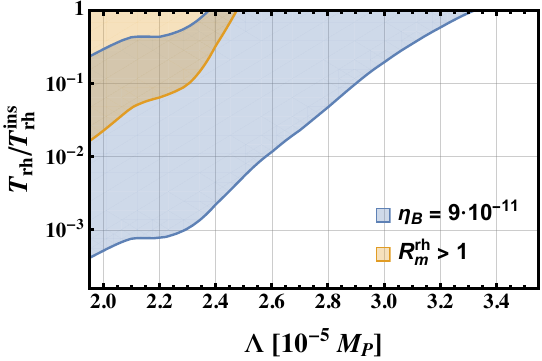}
\caption{Similar to Fig.~\ref{baryogenesis_wo_schwinger}, these figures again display a scan of the parameter $\Lambda$ with the first set of initial conditions, BP$a$ in Tab.~\ref{parmeterchoices} including the Schwinger effect. Top panels: Magnetic and electric energy density (left) and their ratio with the inflation energy density (right). Middle panel: baryon asymmetry $\eta_B$ (left) and Reynolds numbers (right) with their corresponding constraints in red. Bottom panels: CPI temperature with constraint in red (left) and baryogenesis parameter space (right). We did not display the CPI temperature on this last plot as the CPI is no longer a constraint.}
\label{baryogenesis_w_schwinger}
\end{figure}

The production of hypermagnetic fields nevertheless happens at the inflationary scale, hence one must ensure that the helicity is preserved as the Universe cools down in the radiation-dominated era that follows reheating. A rough estimate is to require that the magnetic Reynolds number $\Rm$ is bigger than unity, as this implies that the effects of magnetic induction are dominating over magnetic diffusion in the thermal plasma. On the other hand, the electric Reynolds number $\Re$ determines in which regime the plasma evolves and informs us how to calculate the magnetic Reynolds number, see e.g.~Refs. \cite{Domcke:2019mnd, Cado:2022evn}. In our work, we found that we are in the viscous regime, $\Re < 1$, and hence we need to satisfy the constraint
\begin{align}
\Rm^{\rm rh} \approx 5.9 \cdot 10^{-6} \; \frac{\rho_{B_Y} \ell_{B_Y}^2}{H(t_{\text{end}})^2}  \left( \frac{H(t_{\text{end}})}{10^{13} \, \text{GeV}} \right) \left(\frac{T_{\rm rh}}{T_{\rm rh}^{\rm ins}}  \right)^{\frac{2}{3}} >1,\end{align}
where $\ell_{B_Y}$ is the hypermagnetic characteristic size given by 
\begin{align}\ell_{B_Y}  = \frac{2\pi}{\rho_B\,a^3}  \int_{k_{\text{min}}}^{k_c}d k \, \frac{k^3 }{4\pi^2} \sum_\lambda |u^\lambda|^2.
\end{align}
The magnetohydrodynamics description of the plasma also admits a CP-odd term that can induce a helicity cancellation because of the fermion asymmetry back-transformation into helical gauge fields with opposite sign. This is because the energy configuration in the gauge sector is more favorable than in the fermion sector \cite{Joyce:1997uy}, a phenomenon called chiral plasma instability~(CPI).
Thus, one must ensure that all fermion asymmetry created alongside the helical field during inflation is erased by the action of the weak sphaleron for $10^{12}\text{ GeV}\gtrsim T\gtrsim 130$~GeV. Because the weak interaction only couples to left-handed fermions, the right-handed fermions are protected from the washout until their Yukawa interaction becomes relevant in thermal equilibrium. The right-handed electron $e_R$ is the last species to come into chemical equilibrium, at temperatures $\sim 10^5\,$GeV, thus its asymmetry survives the longest. Therefore, to efficiently erase the fermion asymmetry, while preserving the helicity in the gauge sector, before the CPI can happen, one must require that \cite{Joyce:1997uy, Domcke:2019mnd}
\begin{align} 
T_{\rm CPI}  \approx  (4 \cdot 10^{-7}\;\text{GeV}) \,  \; \frac{\H_Y^2}{H(t_{\text{end}})^6} \, \left( \frac{H(t_{\text{end}})}{10^{13} \, \text{GeV}} \right)^3\left(\frac{T_{\rm rh}}{T_{\rm rh}^{\rm ins}}  \right)^2 \lesssim 10^5\text{ GeV}. \label{constraint-TCPI} 
\end{align}
In Figs.~\ref{baryogenesis_wo_schwinger} and~\ref{baryogenesis_w_schwinger}, we display the main results for the baryogenesis mechanism both in the presence and absence of the Schwinger effect. In both figures, the top panels display the electromagnetic energy and energy ratio to the background energy density. In the middle panels we show the quantities $\eta_B(\Lambda, f_{\theta_W})$, $\mathcal{R}_m(\Lambda)$ and $\mathcal{R}_e(\Lambda)$.
On the left, the red line must be in between the two curves to meet the constraint. On the right, the only constraint is that $\mathcal{R}_m$ is above the red line. Finally, at the bottom, we present the CPI temperature as a function of $\Lambda$ and the regions where the different constraints are met.
On the bottom left panel, the curve should be below the red line. On the right one, we shall seek the overlapping region. In this last plot, we add the temperature ratio $T_{\rm rh}\,/ \, T_{\rm rh}^{\rm ins}$ as a supplementary parameter. We see that the window is larger in the presence of the Schwinger, which also totally removes the constraint on $T_{\rm CPI}$. Indeed, the backreactionless mechanism tends to overshoot the BAU, an issue addressed by the presence of the Schwinger effect which therefore acts as a BAU facilitator.

\section{Summary and Conclusions}\label{disc}
We have discussed baryogenesis in the context of $R^2$-Higgs inflation, involving the CP-violating dimension-six term proportional to $(R/\Lambda^2) \, B_{\mu\nu} \tilde{B}^{\mu\nu}$. We adopt a fully covariant formalism for both inflationary dynamics and gauge field production. Our linear order analysis shows that if $\Lambda\sim 3\times 10^{-5}M_{\rm P}$, sufficient helical hypermagnetic fields are produced, which can lead to the observed BAU during the electroweak crossover. Smaller values of $\Lambda$ imply an overproduction of baryons.
Once the Schwinger effect is included, the energy densities $\rho_E$ and $\rho_B$ are suppressed, but there is a subtlety: the Schwinger effect is exponentially suppressed by a factor $\sim \exp(-\sqrt{(2/3)}\, {\varphi}/{M_{\rm P}})$, which dilutes its relevance during the inflationary epoch, but becomes pronounced around and after the end of inflation. The Schwinger effect can then lead to baryogenesis for smaller values $\Lambda \sim 2.2 \times 10^{-5} M_{\rm{P}}$. We also find that when the Schwinger effect is included, the radiation density $\rho_{\psi}$ can dominate over the electromagnetic densities $\rho_E$ and $\rho_B$, cf.~Fig.~\ref{hubbleenergy_breakdown}. 

We have primarily focused on the Starobinsky-like regime in our linear order analysis. In the mixed $R^2$-Higgs scenario, a smaller $\Lambda$ may generate BAU without the Schwinger effect. This can be understood from Eq.~\eqref{eq:Bmu} where a smaller $\xi_R$ and moderately large $\xi_H$ (i.e.~the mixed $R^2$-Higgs like regime) can induce inflation,
while BAU can be triggered by a larger scale $\Lambda$. However, a larger $\xi_H$ may lead to an exponential growth of isocurvature modes (see e.g.~Refs.~\cite{Bassett:1999ta,Liddle:1999hq,Gordon:2000hv}) in our backreactionless scenario although such a mode is suppressed during inflation. Moreover, in such a scenario, one would need to take into account non-perturbative effects. In our analysis, we have not considered the impact of decay and self-resonance. Thus, the ratio $T_{\rm rh}/T_{\rm rh}^{\rm ins}$ is essentially a free parameter in our analysis. We leave a more detailed analysis of (p)reheating and particle production for future work. It has been pointed out that the helical gauge fields may source non-gaussianity~\cite{Barnaby:2010vf,Barnaby:2011qe}, which may result in moderate constraint to the parameter space for baryogenesis without the Schwinger effect~\cite{Cado:2022evn}. In the presence of the Schwinger effect, the produced helical gauge fields are much weaker and we expect those constraints to be harmless.
However, one needs to be careful when interpreting results from Refs.~\cite{Barnaby:2010vf,Barnaby:2011qe} as they focus on a single field. In our multi-field model, a proper estimation of non-gaussianity requires considering perturbations up to third order. This would induce several new contributions from field-space Riemann tensor~\cite{Kaiser:2012ak} and is beyond the scope of our work.

While there are many avenues to achieve the observed BAU, baryogenesis driven by a dimension-six CP-odd term $\sim (R/\Lambda^2) \, B_{\mu\nu} \tilde{B}^{\mu\nu}$ provides a motivated approach to address BAU within the framework of $R^2$-Higgs inflation. This approach critically rests on the presence of an effective dimension-six term, but it does not require additional degrees of freedom beyond the SM. In parallel, such dimension-six terms can also shed light on the UV sensitivity of $R^2$-Higgs inflation as discussed in, e.g.~Refs.~\cite{Modak:2022gol,Lee:2023wdm}.

\medskip
\textbf{Acknowledgments} --- YC acknowledges funding support from the
Initiative Physique des Infinis (IPI), a research training program of
the Idex SUPER at Sorbonne Universit\'{e}. CE is supported by the UK
Science and Technology Facilities Council (STFC) under grant
ST/X000605/1 and the Leverhulme Trust under RPG-2021-031. TM is funded
by the Deutsche Forschungsgemeinschaft (DFG, German Research
Foundation) under grant 396021762 — TRR 257: Particle Physics
Phenomenology after the Higgs Discovery and Germany’s Excellence
Strategy EXC 2181/1 — 390900948 (the Heidelberg STRUCTURES Excellence
Cluster). The work of MQ is partly supported by Spanish MICIN under
Grant PID2020-115845GB-I00, and by the Catalan Government under Grant
2021SGR00649. IFAE is partially funded by the CERCA program of the
Generalitat de Catalunya.

\appendix

\section{The Vierbein Fields}\label{vier}

The vierbein fields $e^a_\mu$ are defined as follows: The metric in the Jordan frame $g_{J\mu\nu}$ can be related at every point to a Minkowski tangent space $\eta_{ab}$ via the 
vierbein, which obeys the following orthogonality conditions
\begin{align}
e_\mu^a e_a^\nu = \delta^\nu_\mu,~~e_\mu^a e_b^\mu = \delta^a_b,~~g_{J\mu \nu} =  e_\mu^a e_\nu^b \eta_{ab}~\mbox{and}~\gamma_\mu = e^a_\mu \tilde{\gamma}_a,
\end{align}
where $\tilde{\gamma}_a$ are the Minkowski $\gamma$-matrices. The $\gamma_\mu$  satisfy $\{\gamma^\mu,\gamma^\nu\}= 2 g_J^{\mu\nu}$ in curved space-time.
The spin-affine connection is given by 
\begin{equation}
\Gamma_\mu = \frac{1}{2} \omega_{\mu\; ab} \sigma^{ab}~\text{with}~\sigma^{ab} =\frac{1}{4}[\tilde{\gamma}^a, \tilde{\gamma}^b].
\end{equation}
The spin-connection $\omega^{\;a}_{\mu \; \;b}$ is defined as~\cite{Collas:2018jfx} 
\begin{align}
\omega^{\;a}_{\mu \; \;b}\equiv \left( e_{\nu}^a e^\beta_b \Gamma^\nu_{\mu\beta} - e^\beta_b \partial_\mu e_\beta^a \right).
\end{align}

\section{Field-Space Metric and Christoffel Symbols}\label{fieldchris}
The field-space $G^{IJ}$ metric is given by
\begin{align}
G^{\phi\phi} =1,~~G^{hh}= e^{\sqrt{\frac{2}{3}}\frac{\phi}{M_{\rm P}}},~~G^{\phi h} = G^{h \phi} =0.
\end{align}
The corresponding non-vanishing Christoffel symbols are therefore
\begin{align}
\Gamma^{\phi}_{hh}= \frac{e^{-\sqrt{\frac{2}{3}}\frac{\phi }{M_{\rm P}}}}{\sqrt{6}M_{\rm P}},~~\Gamma^{h}_{\phi h} = \Gamma^{h}_{h\phi}= -\frac{1}{\sqrt{6} M_{\rm P}}.
\end{align}

\section{Einstein Equation and Stress-Energy Tensor}\label{energymomtensor}
The action $S_E$ can be rewritten in the following way
\begin{align}
S_E  &= \int d^4 x \sqrt{-g_E}\bigg(\frac{M_{\rm P}^2}{2} R_E +\mathcal{L}_M\bigg),
\end{align}
where $\mathcal{L}_M$ is all terms in the action other than $R_E$. Varying the action with respect to $g_E^{\mu\nu}$
we get 
\begin{align}
0 & = \delta S_E \nn \\
&= \int d^4 x \left( \frac{M_{\rm P}^2}{2} \frac{\delta (\sqrt{-g_E} R_E)}{\delta g_E^{\mu\nu}}+ \frac{\delta (\sqrt{-g_E} \mathcal{L}_M)}{\delta g_E^{\mu\nu}}\right)\delta g_E^{\mu\nu} \\
&=\int d^4 x \left( \frac{M_{\rm P}^2}{2} \left(R_E \frac{\delta  (\sqrt{-g_E} )}{\delta g_E^{\mu\nu}} + \sqrt{-g_E} \frac{\delta  (R_E)}{\delta g_E^{\mu\nu}}\right)+
\sqrt{-g_E} \frac{\delta (\mathcal{L}_M)}{\delta g_E^{\mu\nu}} + \mathcal{L}_M \frac{\delta (\sqrt{-g_E})}{\delta g_E^{\mu\nu}}\right) \delta g_E^{\mu\nu}.\nn
\end{align}
Utilizing $ \frac{\delta  (\sqrt{-g_E} )}{\delta g_E^{\mu\nu}} =  -\frac{1}{2}\sqrt{-g_E}  g_{E\mu\nu}$, $\frac{\delta  (R_E)}{\delta g_E^{\mu\nu}} = R_{E\mu\nu}$ and ignoring 
the surface term we get
\begin{align}
R_{E\mu\nu} -\frac{1}{2} g_{E\mu\nu} R_E &= \frac{1}{M_{\rm P}^2}T_{\mu\nu} 
\end{align}
where 
\begin{align}
T_{\mu\nu} = \left(\mathcal{L}_M g_{E\mu\nu} - 2 \frac{\delta \mathcal{L}_M}{\delta g_E^{\mu\nu}}\right),
\end{align}
which is found to be
\begin{equation}
\begin{split}
\label{eq:energymomten}
T_{\mu\nu} &= \bigg[G_{IJ} D_\mu \phi^I D_\nu \phi^J + g_E^{\alpha\beta} B_{\alpha\mu}B_{\beta\nu}  
+  g_E^{\alpha\beta} W^i_{\alpha\mu}W^i_{\beta\nu} + \frac{1}{2} e^{-\sqrt{\frac{2}{3}}\frac{\phi}{M_{\rm P}}}
g^2 h^2 \frac{(W^1_\mu-i W^2_\mu)}{\sqrt{2}} \frac{(W^1_\nu+i W^2_\nu)}{\sqrt{2}} \\
& +
\frac{2 M_{\rm P}^2}{\xi_R\Lambda^2\sqrt{-g_E}}
F(\phi^I)e^{\sqrt{\frac{2}{3}}\, \frac{\phi}{M_{\rm P}}} \left( 2 g_{E\mu\alpha} \epsilon^{\alpha\beta\rho\sigma} B_{\nu\beta} B_{\rho\sigma} + \frac{1}{8} g_{E\mu\nu} 
\epsilon^{\alpha\beta\rho\sigma} B_{\alpha\beta} B_{\rho\sigma}\right)\\
&+  \frac{1}{4} e^{-\sqrt{\frac{2}{3}}\frac{\phi}{M_{\rm P}}}  h^2 \left(g W^3_\mu - g' B_\mu \right) \left( g W^3_\nu - g' B_\nu \right)
 \bigg] -\\
& g_{E\mu\nu}\bigg(\frac{1}{2} G_{IJ} g_E^{\alpha \beta} D_\alpha \phi^I D_\beta \phi^J + 
V_0(\phi^I) +\frac{2 M_{\rm P}^2}{\xi_R\Lambda^2\sqrt{-g_E}}
F(\phi^I)e^{\sqrt{\frac{2}{3}}\, \frac{\phi}{M_{\rm P}}}B_{\alpha\beta}\tilde B^{\alpha\beta} \\
&+  \dfrac{1}{4} g_E^{\alpha\rho} g_E^{\beta\sigma} B_{\alpha\beta}B_{\rho\sigma}  + \dfrac{1}{4} g_E^{\alpha\rho} g_E^{\beta\sigma} W^i_{\alpha\beta}W^i_{\rho\sigma}
 + \frac{1}{4 } e^{-\sqrt{\frac{2}{3}}\frac{\phi}{M_{\rm P}}} g_E^{\alpha\beta} g^2 h^2 \frac{(W^1_\alpha-i W^2_\alpha)}{\sqrt{2}} \frac{(W^1_\beta+i W^2_\beta)}{\sqrt{2}}\\
 & +  \frac{1}{8 } e^{-\sqrt{\frac{2}{3}}\frac{\phi}{M_{\rm P}}} g_E^{\alpha\beta} h^2 \left(g W^3_\alpha - g' B_\alpha \right) \left( g W^3_\beta - g' B_\beta \right)
 \bigg)+ e^{-\sqrt{\frac{2}{3}}\frac{\phi}{M_{\rm P}}} \bar f e^a_{\mu} \tilde{\gamma}_a \nabla^f_{\nu}  f. 
\end{split}
\end{equation}

\section{The Valley Approximation}\label{valleyapprox}
In this section, we detail aspects of the so-called valley approximation for $V_0$. In this approximation, the system essentially behaves as 
a single-field scenario. Firstly, for positivity of the potential at the inflationary scale, one requires 
\begin{align}
\lambda + \frac{\xi_H^2}{4\xi_R} >0.
\end{align}
For solving the background equations and the inflationary dynamics we focus on the $ R^2$-like regime and the initial condition 
of the valley approximation derives from
\begin{align}
\frac{\partial V_0}{\partial h}=0,
\end{align}
which gives three solutions
\begin{align}
h =0,~~\mbox{and}~~h =\pm\frac{\sqrt{ e^{\sqrt{\frac{2}{3}}\frac{\phi}{M_{\rm P}}}-1}}{\sqrt{4\lambda +\frac{\xi_H^2}{\xi_R}}} \sqrt{\frac{\xi_H}{\xi_R}} M_{\rm P}\label{eq:valley}.
\end{align}
One may choose the trivial solution $h=0$, or the solution with a positive sign for convenience.

\section{Numerical Solutions of the Electromagnetic Equations}\label{rk4method}
In the following, we summarize the details of solving the mode equation of Eq.~\eqref{eq:modephoton} in cosmological time $t$ using the RK4 method
\begin{align}
&\ddot{u}^\lambda + H \dot{u}^\lambda +
\bigg(\frac{k^2}{a^2}  +
\frac{8 \cos^2\theta_W M_{\rm P}^2}{\xi_R \Lambda^2 a} \partial_0\left( F(\varphi^I)e^{\sqrt{\frac{2}{3}}\, \varphi/M_{\rm P}} \right)\lambda k  \bigg)u^\lambda =0.
\end{align}
Firstly, as required for the RK4 method, we rewrite the above equation as two first order equations
\begin{align}
\frac{du^\lambda }{dt} = y^\lambda~~~\mbox{and},~~~
\frac{dy^\lambda}{dt} = - y^\lambda H -\bigg(\frac{k^2}{a^2}  + \frac{8 \cos^2\theta_W M_{\rm P}^2}{\xi_R \Lambda^2 a} \partial_0\left( F(\varphi^I)e^{\sqrt{\frac{2}{3}}\, \varphi/M_{\rm P}} \right)\lambda k  \bigg)u^\lambda
\end{align}
The equations are essentially in the form of 
\begin{align}
\frac{du^\lambda }{dt}= f(u^\lambda, y^\lambda,t),~~~\mbox{and}~~~\frac{dy^\lambda}{dt} = g(u^\lambda, y^\lambda,t),
\end{align}
with
\begin{equation}
\begin{split}
&f(u^\lambda, y^\lambda,t) = y^\lambda,\\
&g(u^\lambda, y^\lambda,t) = - y^\lambda H -\bigg(\frac{k^2}{a^2} 
+ \frac{8\cos^2\theta_W M_{\rm P}^2}{\xi_R \Lambda^2 a} \partial_0\left( F(\varphi^I)e^{\sqrt{\frac{2}{3}}\, \varphi/M_{\rm P}} \right)\lambda k  \bigg)u^\lambda.
\end{split}
\end{equation}
Now the task is to find out $u^\lambda$ and $y$ for each time step utilizing the RK4 method. This is provided by
\begin{equation}
\begin{split}
&u^\lambda_{i+1} = u^\lambda_i + \frac{1}{6}\left(l_0 + 2 l_1 + 2l_2 +l_3\right),\\
&y^\lambda_{i+1} = y^\lambda_i + \frac{1}{6}\left(m_0 + 2 m_1 + 2m_2 +m_3\right),
\end{split}
\end{equation}
with
\begin{equation}
\begin{split}
&l_0 = \delta t~ f(u^\lambda_i, y^\lambda_i,t_i),\\
&m_0 = \delta t~ g(u^\lambda_i, y^\lambda_i,t_i),\\
&l_1 = \delta t~ f(u^\lambda_i +\frac{1}{2} l_0, y^\lambda_i+\frac{1}{2} m_0,t_i + \frac{1}{2} \delta t),\\
&m_1 = \delta t~ g(u^\lambda_i +\frac{1}{2} l_0, y^\lambda_i+\frac{1}{2} m_0,t_i + \frac{1}{2} \delta t),\\
&l_2 = \delta t~ f(u^\lambda_i +\frac{1}{2} l_1, y^\lambda_i+\frac{1}{2} m_1,t_i + \frac{1}{2} \delta t),\\
&m_2 = \delta t~ g(u^\lambda_i +\frac{1}{2} l_1, y^\lambda_i+\frac{1}{2} m_1,t_i + \frac{1}{2} \delta t),\\
&l_3 = \delta t~ f(u^\lambda_i + l_2, y_i+ m_2,t_i ),\\
&m_3 = \delta t~ g(u^\lambda_i + l_2, y_i+ m_3,t_i),
\end{split}
\end{equation}
where $\delta t$ is the time step. The Bunch-Davis initial conditions for the modes $u^\lambda$ and $y$ are given in Eq.~\eqref{eq:ini}.

One can in principle fix the number of modes $N_k$ in each time step within $\left[k_{\rm{min}}, k_c\right]$ for the integration of Eqs.~\eqref{eq:plasma-observables}. However, this makes the initialization of the modes in the next time step more involved. This is because as $k_c$ increases in each time step, keeping $N_k$ fixed each time would require some more involved initialization for subsequent time steps. We can take a simpler route and keep the number of $k$ modes the same for all time steps. This ensures that the number of modes $N_k$ and the corresponding modes are identical at each time step. In practice, we take a large range $\left[k_{\rm{min}}, k_{\rm{max}}\right]$ with $k_{\rm{max}} = C a(t_{\rm{numend}}) H(t_{\rm{numend}})$ where $t_{\rm{numend}}$ is the numerical end of our simulation. We chose $C=100$ to ensure that $k_c(t_{\rm{numend}}) <  k_{\rm{max}}$ and divide the range $\left[k_{\rm{min}}, k_{\rm{max}}\right]$ into $N_k =200$ intervals. In each time step, we then numerically interpolate Eqs.~\eqref{eq:plasma-observables} in $\left[k_{\rm{min}}, k_{\rm{max}}\right]$ and truncate the numerical integration up to the corresponding $k_c$ values. Increasing $N_k$ to higher values does not significantly impact our results. For further details of the numerical procedure, we refer the reader to Ref.~\cite{Cado:2022pxk}.

In the presence of the Schwinger effect the corresponding equation of motion, Eq.~(\ref{eq:modephoton_sch}), is solved numerically using similar methods as those described above.

\bibliography{references}

\end{document}